\documentclass[bibyear]{aa}     % if the references are not structured
\usepackage{natbib}
\usepackage{graphicx}
\usepackage{txfonts}
\usepackage{lipsum}
\usepackage{subcaption}         % necessary for continued figures
\usepackage{lscape}             % to rotate a single page table, example in appendix.
\usepackage{placeins}           % useful with \FloatBarrier, to keep 
\usepackage{hyperref}
\newcommand{\FIRE}{\text{FIREbox$^{HR}$}\xspace}
\newcommand{\Mstar}{\log{(M_{\star}/M_{\odot})}\xspace}
\newcommand{\Msun}{M_{\odot}}
\newcommand{\logit}{\mathop{\mathrm{logit}}\nolimits}
\newcommand{\ETa}{\mathrm{ETlogA}\xspace}
\newcommand{\ltETa}{\mathrm{ltETlogA}\xspace}
\newcommand{\linkFilters}{\href{https://jwst-docs.stsci.edu/jwst-near-infrared-camera/nircam-instrumentation/nircam-filters\#gsc.tab=0}{See NIRCam Filters Documentation}}

\newcommand{\linkPSFs}{\href{https://jwst-docs.stsci.edu/jwst-near-infrared-camera/nircam-performance/nircam-point-spread-functions\#gsc.tab=0}{See NIRCam PSF Documentation}}
\begin{document}

   \title{Inferring the shapes of low-mass galaxies at $z>6$ from \FIRE}

   \subtitle{Where are the disks?}

    \titlerunning{Inferring the shapes of low-mass galaxies at $z>6$ from \FIRE}
    \authorrunning{Geng, Feldmann et al.}

%%%%%%%%%%%%%%%%%%%%%%%%%%%%%%%%%%%%%%%%
% Please do not include ORCIDs next to author names.
% Only ORCIDs authenticated by individual authors in EDP Sciences editorial system will be taken into account.
% ORCIDs included here will be removed.
%%%%%%%%%%%%%%%%%%%%%%%%%%%%%%%%%%%%%%%%

    \author{
    Haojie Geng\inst{1},
    Robert Feldmann\inst{1}, 
    Jorge Moreno\inst{2},
    Xuejian Shen\inst{3}\textsuperscript{,}\inst{4},
    and James Bullock\inst{5},
    }

   \institute{
   \inst{1}\,\textit{Institute of Astrophysics, University of Zurich, Zurich CH-8057, Switzerland} \\
   \inst{2}\,\textit{Department of Physics and Astronomy, Pomona College, Claremont, CA 91711, USA} \\
   \inst{3}\,\textit{Department of Physics, Kavli Institute for Astrophysics and Space Research, Massachusetts Institute of Technology, Cambridge, MA 02139, USA} \\
   \inst{4}\,\textit{Center for Astrophysics $|$ Harvard $\&$ Smithsonian, 60 Garden Street, Cambridge, MA 02138, USA} \\
   \inst{5}\,\textit{Department of Physics and Astronomy, University of Southern California, Los Angeles, CA 90089, USA} \\
   }

   \date{Submitted to A\&A on October 18, 2026}

% \abstract{}{}{}{}{}
% 5 {} token are mandatory
  \abstract
   {The three-dimensional (3D) shapes of galaxies encode key information about their formation and evolution. A forward-modeling Bayesian approach allows us to infer intrinsic shapes from projected half-light radii and axis ratios but the results may depend sensitively on the assumed population-level distribution of intrinsic shape parameters.}
   {We wish to test the validity of commonly adopted distributional assumptions, as well as other modeling details in recent 3D shape inference studies of high-redshift galaxies.}
   {Using the \FIRE simulation, we study galaxies with $\log(M_\star/M_\odot)=6.5-9.5$ over $z\sim6-15$. We measure intrinsic 3D shapes directly from stellar particle distributions with a robust minimum covariance determinant (MCD) estimator. In parallel, we generate mock JWST/NIRCam observations, measure half-light radii and axis ratios, and apply forward-modeling Bayesian inference to recover the intrinsic shape distribution from the projected $(q,\log a)$ joint distribution. 
   %The results from two channels provide a direct constraint on the better distributional assumption.
   }
   {The simulated low-mass galaxies are predominantly prolate with a substantial spheroidal component.
   The intrinsic ellipticity $E$ and triaxiality $T$ distributions are well described by JohnsonSB forms, while intrinsic sizes $A$ are approximately log-normal. 
   Motivated by this, we introduce a logit-transformed population model in which the transformed parameters $(\logit{E},\logit{T},\log A)$ follow a multivariate normal distribution, with general correlations. 
   This model reproduces the particle-measured intrinsic distributions more faithfully than the standard multivariate normal (in $E, T, \log{}A$) formulation. 
   Other assumptions of the Bayesian inference approach are also discussed. Generally, we find that triaxiality-related parameters require large samples or informative priors to be robustly constrained from projected data alone.}
  % conclusions heading (optional)
   {Forward-modeling Bayesian inference is a powerful tool to recover intrinsic 3D shape distributions when the population model matches the true intrinsic parameter distribution. Population-appropriate distribution families are therefore essential to avoid biased structural inferences and misinterpretation of the morphological evolution of galaxies at high redshift.}


   \maketitle
   \nolinenumbers  % disabling linenumbers

%%%%%%%%%%%%%%%%%%%%%%%%%%%%%%%%%%%%%%%%%%%%%%%%%%%%%%%%%%%%%%
\section{Introduction}
\label{section 1: Introduction}
    Over the past several decades, numerous studies have inferred intrinsic shapes from projected morphologies across a wide range of stellar masses and redshifts ($\Mstar = 9$ -- $11$, $z = 0$ -- 8), using imaging from the \textit{Sloan Digital Sky Survey} (SDSS), the \textit{Hubble Space Telescope} (HST), and more recently the \textit{James Webb Space Telescope} (JWST).
    These works typically approximate galaxies as members of a triaxial ellipsoid family that spans three limiting configurations -- prolate (cigar-like), oblate (disk-like), and spheroidal.
    Such ellipsoids are defined by intrinsic axis lengths $A \ge B \ge C$, with corresponding shape parameters expressed through the axis ratios $(B/A,C/A)$.
    
    Early analyses modeled the distribution of projected axis ratios to reconstruct the underlying intrinsic distributions \citep{lambas1992, Yuma2012, Mendez2010, VanderWel2014, Costantin2018}.
    More recently, \citet{Zhang2019} and \citet{Pandya2024} have jointly modeled the distributions of projected axis ratios and sizes, incorporating correlations between intrinsic ellipticity and size that are required to reproduce the characteristic ``banana'' locus observed in the projected axis ratio and half-light radius $(b/a,R_e)$ plane (see Figure~5 of \citealt{Pandya2024}).
    
    Current observational constraints on intrinsic shapes can be broadly divided into the following four regimes: intermediate-redshift ($z$=1.5 -- 2.5) and low-redshift ($z<1.5$) massive galaxies ($\Mstar>10$), and their low mass counterparts ($\Mstar<10$).
    At very early times ($z \gtrsim 2.5$), robust inferences are available primarily from the comprehensive Bayesian analysis of JWST data presented by \citet{Pandya2024}.
    For low-redshift massive galaxies, it is now well established that the population consists predominantly of randomly oriented oblate and spheroidal systems \citep{lambas1992, Alam2002, Ryden2004, Vincent2005, padilla2008, vanderWel2009, Mendez2010, VanderWel2014, Costantin2018, Zhang2019, Pandya2024}.
    At intermediate redshift, massive systems comprise all three shape classes, with the fraction of oblate galaxies increasing toward lower redshift and that of spheroidal galaxies increasing at higher redshift, while prolate systems remain sub-dominant.
    At $z>2.5$, JWST analyses suggest that the oblate component among massive galaxies becomes more prominent and that the scatter in ellipticity decreases, though current conclusions rely on a single study and will need to be revisited with larger samples \citep{Pandya2024}.
    The picture is more complex for dwarf galaxies: multiple studies indicate a transition from predominantly oblate systems at low redshift to primarily prolate or strongly triaxial systems at higher redshifts, accompanied by a tightening of the ellipticity distribution up to $z\sim 8$ \citep{Ferguson2004, Ravindranath2006, Law2012, VanderWel2014, Zhang2019, Pandya2024}.
    
    Cosmological hydrodynamical simulations offer an essential complement to observational work by providing direct access to intrinsic 3D galaxy shapes.
    State-of-the-art simulations, including TNG50 \citep{Pillepich2019}, the VELA zoom-in suite \citep{Zhang2019}, TNG100 \citep{Zhang2022}, and FIREbox \citep{Hopkins2015, klein2026, Xia2026}, broadly reproduce the observed trends for massive galaxies: more massive galaxies are more likely to be oblate/spheroidal. 
    
    However, predictions for dwarf galaxies vary significantly between simulation suites.
    For example, TNG50 finds that low-mass galaxies at high redshift are frequently spheroidal or prolate, with an increasing fraction of oblate systems toward lower redshifts for $\Mstar = 9$ -- $11$, and an increasing spheroidal fraction for $\Mstar = 7$ -- $9$ \citep{Pillepich2019}.
    VELA shows similar behavior \citep{Zhang2019}.
    In comparison, recent FIREbox analysis also indicates that, at $z=0$, galaxies with $\Mstar=8$ -- $10$ are predominantly spheroidal, with a deficit of very thin, low-mass disks relative to local axis-ratio statistics \citep{klein2026}.
    Taken at face value, such results appear in tension with the high fraction of flat galaxies inferred for low-mass galaxies at low $z$ \citep{Benavides2025, Benavides2026} and the oblate-prolate mixture at high $z$ \citep{Pandya2024}.
    
    The apparent discrepancies between observationally inferred and simulated 3D shapes motivate us to revisit the Bayesian framework introduced by \citet{Zhang2019} and \citet{Pandya2024}.
    In both studies, the projected distributions of axis ratio $b/a$ and half-light radius $R_e$ are modeled by assuming that the intrinsic ellipticity, triaxiality, and intrinsic size parameters follow a trivariate Gaussian distribution.
    The general correlation matrix should be considered in principle; however, both studies found that the results are similar when only the correlation between intrinsic ellipticity and size is considered. 
    It is also assumed that a single distribution is enough to describe the whole dataset.
    Both assumptions may or may not be broken when we move our attention to the high-redshift galaxies.
    
    In this paper, we use the \FIRE cosmological simulations to measure intrinsic 3D shapes and corresponding projected properties, with the goal of directly testing the robustness of the current Bayesian inference framework for high-redshift dwarf galaxies.
    We quantify the performance of the inference method by comparing intrinsic shapes obtained through two complementary approaches:
    (i) \emph{inference}, where JWST-like mock observations are generated, projected axis ratios and sizes are measured in different stellar-mass and redshift bins, and the Bayesian formalism of \citet{Zhang2019} and \citet{Pandya2024} is applied; and
    (ii) \emph{direct measurement}, where true 3D shapes are computed from the spatial distribution of star particles.
    We study the dwarf galaxies with $\Mstar = 6.5$ -- $9.5$ at $z = 6.3-15$ while focusing on $z=6-8$, enabling comparisons both to existing observational inferences and to other simulation suites.
    We show that, when supplemented with additional directly measurable quantities, intrinsic shapes can be recovered more faithfully, and that while the current trivariate Gaussian model performs reasonably well in certain regimes, a more flexible intrinsic shape distribution is required to capture the geometric diversity of dwarf galaxies at $z>6$.
    
    This paper is organized as follows.
    Section~\ref{section 2: Data and Method} describes the \FIRE simulations, the construction of mock JWST observations, and our two shape-measurement channels.
    Section~\ref{section 3: Results} presents our findings on intrinsic and projected shapes.
    Section~\ref{section 4: Discussion} discusses the implications of our results and their caveats.
    Finally, Section~\ref{section 5: Conclusion} summarizes our conclusions.
    Throughout this work we assume a \citet{Planck2016} cosmology: $h=0.6774$, $\Omega_{m,0}=0.3089$, $\Omega_{\Lambda,0}=0.691$, and $\Omega_{b,0}=0.0486$.
    Projected and intrinsic shape parameters are denoted $(b/a,\log a)$ and $(B/A,C/A,A)$, respectively, with alternative intrinsic parameterizations defined as ellipticity $E = 1 - C/A$ and triaxiality $T = \bigl[1-(B/A)^2\bigr]/\bigl[1-(C/A)^2\bigr]$.
%%%%%%%%%%%%%%%%%%%%%%%%%%%%%%%%%%%%%%%%%%%%%%%%%%%%%%%%%%%%%%
\section{Data and Method} % mock observation and method
\label{section 2: Data and Method}

\begin{figure*}
    \centering
    \includegraphics[width=1\linewidth]{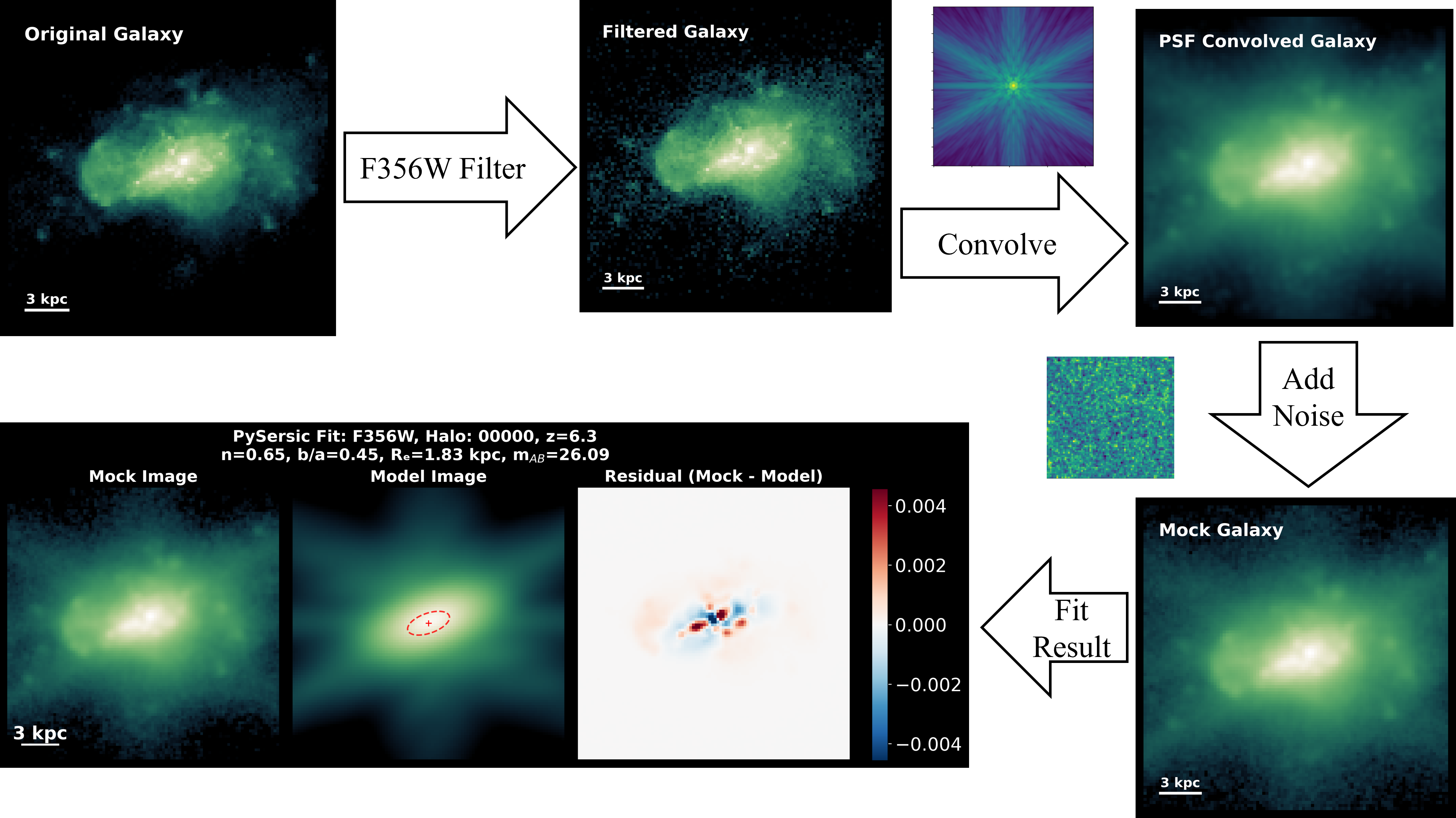}
    \caption{Example of the mock-observation generation pipeline. Starting from the \texttt{SKIRT} output data cube (shown as the image integrated over all wavelengths), we filter the emission through the NIRCam F356W throughput, convolve with the F356W simulated PSF, add a drizzle-correlated background noise map, and finally perform a single-component S\'ersic fit using \texttt{PySersic}. 
    The S\'ersic-fitted galaxy is presented as a residual plot, where on the very left is the mock observation, in the middle is the modeled image overplotted with a centroid red cross and a red ellipse constructed from the s\'ersic parameters, on the right is the residual between the two. 
    All figures are clipped from $10^{-4}*I_{max}$ to $I_{max}$, where $I_{max}$ is the maximum pixel value, and are shown in log scale to present the PSF spikes.
    Fitted s\'ersic index $n$, axis-ratio $b/a$, half-light radius $R_e$ and total flux density converted to apparent AB magnitude $m_{AB}$ are displayed under the title.
    The noise added in this example is reduced by a factor of 1000 compared to the corresponding UNCOVER mosaic so that the true projected shape is recovered with very high signal-to-noise.}
    \label{Mock observation pipeline}
\end{figure*}

\subsection{\texorpdfstring{\FIRE}{FIREboxHR} Simulation}
\label{section 2.1: FIREboxHR Simulation}
    In this work, we analyze the data from the \FIRE cosmological simulation \citep{Feldmann2025}. This simulation is a high-resolution re-run of the original FIREbox project \citep{Feldmann2023}, adopting identical physical prescriptions \citep{Hopkins2018} and the same initial conditions but with increased numerical resolution. We briefly summarize the key properties of the simulation here, referring the reader to \citet{Feldmann2023} and \citet{Feldmann2025} for a comprehensive description of the numerical setup and sub-grid models.

    \FIRE evolves gas, stars, and dark matter within a periodic cubic volume of length $L=22.1$ cMpc, from an initial redshift of $z_{\mathrm{init}}=120$ down to $z=6.3$. The simulation adopts a $\Lambda$CDM cosmology consistent with \citet{Planck2016}. The initial conditions consist of $2\times2048^3$ particles, split equally between gas and dark matter, corresponding to a mass resolution of $m_{\mathrm{gas}} \approx 7823\ \Msun$ and $m_{\mathrm{dm}} \approx 41899\ \Msun$. The gravitational and hydrodynamical evolution of these components is followed using the \texttt{GIZMO} code \citep{Hopkins2015} in its meshless finite-mass (MFM) mode.

    The simulation incorporates realistic baryonic physics and feedback processes via the FIRE-2 model \citep{Hopkins2018}. Star formation occurs only in dense ($n>1000\,{\rm cm}^{-3}$), self-gravitating, Jeans-unstable gas, with a local star formation efficiency set to $\rho_\mathrm{SFR}=f_\mathrm{H2}\,\rho_\mathrm{gas}/t_\mathrm{ff}$, where $t_\mathrm{ff}$ is the local free-fall time and $f_\mathrm{H2}$ is the molecular gas fraction. The model includes explicit treatments for cooling and heating rates arising from free-free emission, photoionization/recombination, metal-line cooling, and molecular processes. Stellar feedback mechanisms, including radiation pressure, photoionization, photoelectric heating, and supernovae (Types Ia and II), are modeled directly based on stellar population synthesis models \citep{Leitherer1999}.

    Dark matter halos and their hosted galaxies are identified using the \texttt{AMIGA} Halo Finder (AHF; \citealt{Knollmann2009}). Halo properties, such as virial mass and radius, are defined using the virial overdensity criterion of \citet{Bryan1998}. We analyze halos containing at least 300 dark matter particles ($M_\mathrm{halo}\gtrsim 10^7\Msun$). This selection yields a sample size of approximately $2.6\times10^4$ halos at $z=15$, increasing to $4.4\times10^5$ halos by $z=6.3$. For the analysis presented in this work, we specifically select the 300 most massive central galaxies at a series of snapshots spanning the redshift range $z=6.3$ to $z=15$ ($z \in \{6.3, 7, 8, 9, 10, 10.8, 11.3, 12, 13.5, 15\}$).

    To facilitate direct comparison with observations, we generate clean mock images using the radiative transfer code \texttt{SKIRT} (version 9; \citealt{Camps2020}). The radiative transfer calculation employs the Binary Population and Spectral Synthesis (BPASS; \citealt{Eldridge2009}) library version 2.2.1 \citep{Stanway2018}, assuming a \citet{Chabrier2003} stellar initial mass function (IMF) with an exponential cutoff below $1\ \Msun$, a high-mass slope of $-2.3$, and a maximum stellar mass of $300\ \Msun$. The dust-to-metal ratio is set to 0.4. The output of this pipeline is a spectral data cube storing the quantity $\lambda I_\lambda(x,y)$ in units ${\rm W}/{\rm m}^2/{\rm arcsec}^2$ where $I_\lambda(x,y)$ is the specific intensity, representing the spectral surface brightness at projected spatial position $(x,y)$ and wavelength $\lambda$. Further details on the mock observation pipeline are provided in \citet{Feldmann2025}.

\subsection{Mock Observation}
\label{section 2.2: Mock Observation}

\urldef{\urlFilters}\url{https://jwst-docs.stsci.edu/jwst-near-infrared-camera/nircam-instrumentation/nircam-filters#gsc.tab=0}

\urldef{\urlPSFs}\url{https://jwst-docs.stsci.edu/jwst-near-infrared-camera/nircam-performance/nircam-point-spread-functions#gsc.tab=0}

    Starting from the \texttt{SKIRT} spectral cubes $\lambda I_\lambda(x,y,\lambda)$, we construct JWST-like mock observations by applying the relevant instrumental and observational effects.  
    We adopt the NIRCam throughputs\footnote{\linkFilters} and corresponding simulated point-spread functions\footnote{\linkPSFs} (PSFs) in six broad bands: F115W, F150W, F200W, F277W, F356W, and F444W.
    For each filter, the filter-integrated spectral surface-brightness density $\bar{I}(x,y)$ is computed as
    \begin{equation}
        \bar{I}(x,y) =
        \frac{\displaystyle \int \frac{{\rm d}\lambda}{c}\,\lambda\,n(\lambda)\,I_\lambda(x,y,\lambda)}
             {\displaystyle \int {\rm d}\lambda\,n(\lambda)/\lambda},
        \label{Eq 1: filter-averaged surface brightness}
    \end{equation}
    where $c$ is the speed of light and $n(\lambda)$ is the dimensionless transmission curve of the filter.  
    The resulting $\bar{I}(x,y)$ map is convolved with the corresponding NIRCam PSF kernel to mimic the signal recorded on the detector.
    
    To reproduce the drizzle-correlated background noise of real JWST mosaics, we use the UNCOVER survey \citep{Bezanson2024} as a template.  
    For each of the six NIRCam bands, we extract a small, source-free patch from the public mosaics and use it to characterize the noise properties.  
    We first estimate the noise amplitude as the standard deviation of pixel values about zero and then compute the two-dimensional autocovariance function (ACF) of the patch to capture the spatial correlations introduced by the reduction pipeline.  
    A discrete Fourier transform of the ACF yields the corresponding power spectral density (PSD).  
    In Fourier space, an uncorrelated white-noise realization $n_{\mathrm{uncorr}}(k)$ drawn from a normal distribution $\mathcal{N}(0,1)$ is converted into a correlated noise field via
    \begin{equation}
        n_{\mathrm{corr}}(k) = n_{\mathrm{uncorr}}(k)\,\sqrt{\mathrm{PSD}(k)}.
    \end{equation}
    The real-space correlated noise map $n_{\mathrm{corr}}(x,y)$ is then obtained by an inverse Fourier transform and rescaled such that its standard deviation matches that of the original source-free patch.

    \begin{figure*}
        \includegraphics[width=\linewidth]{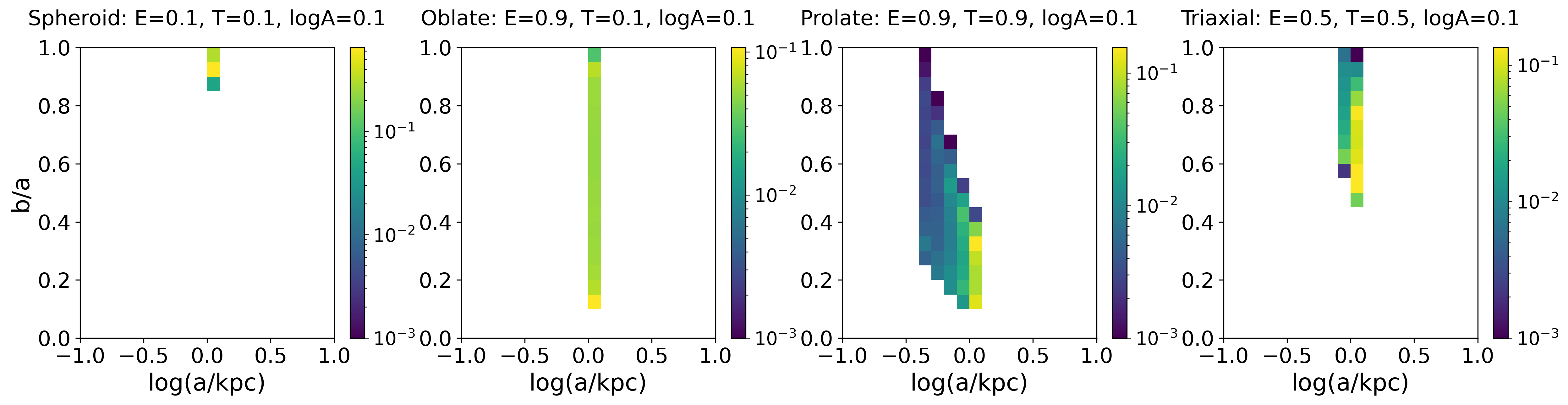}
        \caption{Representative elements of the ellipsoid library.  
            Each panel shows the probability distribution of projected axis ratio $q=b/a$ and projected major-axis size $\log(a/{\rm kpc})$ for a single intrinsic shape $\vec{\theta} = (E,T,\log A)$, obtained from $10^5$ random viewing angles.  
            From left to right: a nearly spheroidal ellipsoid $(E=0.1,T=0.1,\log A=0.1)$, an oblate system $(E=0.9,T=0.1,\log A=0.1)$, a prolate system $(E=0.9,T=0.9,\log A=0.1)$, and a triaxial system $(E=0.5,T=0.5,\log A=0.1)$.  
            Spheroids are always observed as nearly round; oblate systems populate a wide range of $q$ at fixed size; prolate systems produce the characteristic “banana-shaped’’ locus due to their preference for edge-on projections; and triaxial systems populate intermediate regions of the $(q,\log a)$ plane.}
        \label{Ellipsoid library}
    \end{figure*}
    
    The outputs at each stage of the pipeline are images with a pixel scale of $0.04\arcsec\,\mathrm{pix}^{-1}$, matching the UNCOVER mosaics.  
    For each simulated galaxy, we generate mock images for three orthogonal projections of the \texttt{SKIRT} cube, which we treat as statistically independent realizations.  
    Visual inspection of the mock images with full UNCOVER-like noise indicates that more than half of the simulated galaxies are effectively undetectable at that depth.  
    Because our goal is to test the shape-inference methodology rather than to model survey completeness, we therefore increase the signal-to-noise ratio by a factor of $10^{3}$. This is done by scaling down the correlated noise maps by a factor of $10^{3}$ in amplitude relative to the UNCOVER level, corresponding to an effective exposure time larger by a factor of $10^{6}$ under background-limited conditions.  
    This configuration yields very high signal-to-noise ratios (S/N) for the galaxies of interest.  
    For galaxies with $\mathrm{S/N} \gtrsim 10$, the best-fit projected shapes are expected to be only weakly biased by noise, so further increases in S/N primarily reduce the statistical uncertainties rather than systematically altering the inferred axis ratios and sizes.  
    In the following, we make primary use of the F356W and F444W mock observations, which probe rest-frame optical emission up to $z \sim 10$.
    
    To measure galaxy sizes and projected axis ratios, we first use the \texttt{photutils} Python package \citep{Bradley2024} for background estimation and subtraction, source detection, segmentation, and deblending using the settings listed in Table \ref{Source Detection parameters}.
    \begin{table}[!ht]
        \centering
        \caption{Source Detection Parameters}
        \label{Source Detection parameters}
        \begin{tabular}{lc}
        \hline\hline
        Name                    &  Value       \\
        \hline
        background-cell-size    &  50 pixels   \\
        smoothing-box-size      &  3  pixels   \\
        detection-threshold     &  5 $\sigma$  \\
        detect-minarea          &  5-10 pixels \\
        partition-min-contrast  &  0.1         \\
        \hline
        \end{tabular}
    \end{table}
    
    We then employ \texttt{PySersic} \citep{Pasha2023} to fit single-component \citet{Sersic1963} models to the target galaxies, providing the appropriate NIRCam PSF so that the model is convolved to the data resolution.  
    The S\'ersic surface-brightness profile is
    \begin{equation}
        I(r) = I_e \exp\left\{-b_n\left[\left(\frac{r}{R_e}\right)^{1/n} - 1\right]\right\},
        \label{Eq 2: Sérsic profile}
    \end{equation}
    where $r$ is the elliptical radius defined by a position angle and minor-to-major axis ratio $q \equiv b/a$, $I_e$ is the surface brightness at the effective radius $R_e$, $n$ is the S\'ersic index, and $b_n$ is a constant chosen such that $R_e$ encloses half of the total model flux (the half-light radius).  
    All sources other than the central target are masked prior to fitting.  
    Unless stated otherwise, we adopt the default priors provided by \texttt{PySersic} and take the maximum-a-posteriori (MAP) estimate obtained via Laplace’s approximation as the best-fitting set of structural parameters.

\subsection{Bayesian Inference of Galaxy Shapes}
\label{section 2.3: Inferring 3D Shapes}
    Following \citet{Zhang2019} and \citet{Pandya2024}, we approximate galaxies as triaxial ellipsoids whose intrinsic 3D shapes are fully specified by their principal axis lengths $(A,B,C)$ with $A \geq B \geq C$.  
    Equivalently, one may work with the intrinsic axis ratios $(B/A, C/A)$ and a size parameter $A$, or with the “ellipticity-triaxiality-size” parameters
    \begin{equation}
        \vec{\theta} = (E, T, \log A),
        \label{Eq 4}
    \end{equation}
    where $E$ and $T$ denote intrinsic ellipticity and triaxiality as defined in Sect.~\ref{section 1: Introduction}.  
    In this parametrization, oblate systems have $E \simeq 1$ and $T \simeq 0$, prolate systems have $E \simeq 1$ and $T \simeq 1$, and spheroidal systems occupy the region of low ellipticity, $E \sim 0$.
    
    The division between oblate, prolate, and spheroidal shapes is not unique when the parameters do not lie close to these limiting cases.  
    We follow the classification scheme shown in Fig.~1 of \citet{Zhang2019} (see also Fig.~2 of \citealt{VanderWel2014}, Fig.~1 of \citealt{Xia2026}, and \citealt{klein2026}), which partitions the $(C/A,B/A)$ plane into three regions corresponding to spheroidal, oblate, and prolate ellipsoids.
    
    For a given intrinsic shape $(E,T,\log A)$, viewing the ellipsoid from a direction specified by polar angle $\vartheta$ and azimuthal angle $\phi$ yields an ellipse on the sky with semi-axes $a \geq b$.  
    Following the projection formalism of \citet{Binney1985} or Eqs.~(5)-(7) of \citet{vandeVen2021}, we compute the projected axis ratio
    \begin{equation}
        q \equiv \frac{b}{a}
        \label{Eq 5}
    \end{equation}
    and the corresponding projected major-axis size $\log a$ for any orientation $(\vartheta,\phi)$.
    
    \subsubsection{Ellipsoid library in the \texorpdfstring{$(b/a,\log a)$}{(b/a, log a)} plane}
    
    For each intrinsic shape vector $\vec{\theta} = (E,T,\log A)$, we generate a large set of random orientations $(\cos\vartheta,\phi)$, with $\cos\vartheta$ sampled uniformly in $[-1,1]$ and $\phi$ uniformly in $[0,2\pi)$.  
    In this work we use $10^5$ random orientations per $\vec{\theta}$.  
    From these we construct a discrete probability map in the observed plane, namely the 2D histogram of $(q,\log a)$, which we denote by
    \begin{equation}
        H(q_k, \log a_k \mid \vec{\theta}),
        \label{Eq 6}
    \end{equation}
    where $k$ indexes the 2D bins.  
    We pre-compute an “ellipsoid library’’ of such maps for $\vec{\theta}$ values on a regular grid, with $E$ and $T$ spanning $[0,1]$ and $\log A$/kpc spanning $[-1,1]$.  
    
    In contrast to \citet{Zhang2019} and \citet{Pandya2024}, we effectively ignore observational uncertainties in $q$ and $\log a$: the extremely low background noise adopted in our mock images (Sect.~\ref{section 2.2: Mock Observation}) implies that measurement errors are negligible for the purposes of this inference.  
    For consistency with the formalism of previous work, we retain the option to scatter the true projected $q$ using a Rice distribution \citep{Rix1995} and propagate that scatter into $\log a$, but in the analysis presented here the corresponding noise amplitudes are set so small ($\Delta q = 10^{-6}$, $\Delta \log a = 0$) that the measured values are effectively equal to the true projected ones.
    
    Figure~\ref{Ellipsoid library} illustrates representative probability maps from this library.  
    Spheroidal systems appear nearly round from any viewing direction and thus populate the upper region of the $(q,\log a)$ histogram.  
    Oblate ellipsoids are equally likely to be observed in round “face-on’’ or thin “edge-on’’ configurations; a perfectly thin disk would have a projected size that is independent of viewing angle and therefore would be distributed nearly uniformly in $q$ at fixed $\log a$.  
    Prolate systems are more likely to be observed in nearly “edge-on’’ configurations, producing the characteristic banana-shaped locus in the $(q,\log a)$ plane discussed by \citet{Pandya2024}.  
    In mixed populations, this prolate component is the primary source of correlation between $q$ and $\log a$.
    
    Given this library, the model 2D histogram of projected $(q,\log a)$ for a population of galaxies with an intrinsic distribution $f(\vec{\theta}\mid \vec{\mu},\Sigma)$ is approximated by a probability-weighted sum over the library elements:
    \begin{equation}
        \lambda_k(\vec{\mu},\Sigma)
        \;=\;
        N \sum_j f(\vec{\theta}_j \mid \vec{\mu},\Sigma)\,
               H(q_k,\log a_k \mid \vec{\theta}_j)\,\Delta V_j,
        \label{Eq 7: model-hist}
    \end{equation}
    where $N$ is the number of galaxies in the sample, $\vec{\theta}_j$ runs over the grid points in the library, $f(\vec{\theta}\mid\vec{\mu},\Sigma)$ is the intrinsic probability density specified below, and $\Delta V_j$ is the volume element around $\vec{\theta}_j$.  
    The $\lambda_k$ are the model-predicted mean counts in bin $k$ as function of the population parameters $\vec{\mu}$ and $\Sigma$.

%%%%%%%%%%%%%%%%%%%%%%%%%%%%%%%%%%%%%%%%%%%%%%%%%%%%%%%%%%%%%%%%%%%%%%%%%%%%%%%%%%%%%%%%%%%
    \subsubsection{Intrinsic shape distribution and logit transform}
    
    \citet{Zhang2019} and \citet{Pandya2024} assume that each of the $N$ observed galaxies is associated with an intrinsic 3D shape vector $\vec{\theta}_i = (E_i,T_i,\log A_i)$, and that these vectors follow a trivariate normal distribution with a mean vector and covariance matrix in $(E,T,\log A)$ space.  
    In their implementation, only the covariance between ellipticity and size is allowed to be non-zero.
    
    Motivated by the distribution of directly measured intrinsic shapes in our simulation (Sect.~\ref{section 3.2: 3D intrinsic shapes}), we instead work with a transformed parameter vector
    \begin{equation}
        \vec{\theta} = (\logit E,\, \logit T,\, \log A)\equiv (\tilde{E}, \tilde{T}, \log{A}),
        \label{Eq 8}
    \end{equation}
    where $\logit x \equiv \ln[x/(1-x)]$ maps the bounded interval $x \in (0,1)$ to the real line.  
    We then assume that $\vec{\gamma}$ follows a trivariate normal distribution with unknown mean vector
    \begin{equation}
        \vec{\mu} = (\mu_{\tilde{E}}, \mu_{\tilde{T}}, \mu_{\log A})
        \label{Eq 9: mean vector}
    \end{equation}
    and an unknown covariance matrix
    \begin{equation}
        \Sigma =
          \begin{bmatrix}
            \sigma_{\tilde{E}}^2
                & \rho_{\tilde{E}\tilde{T}}\,\sigma_{\tilde{E}}\,\sigma_{\tilde{T}}
                & \rho_{\tilde{E}L}\,\sigma_{\tilde{E}}\,\sigma_{L} \\
            \rho_{\tilde{E}\tilde{T}}\,\sigma_{\tilde{E}}\,\sigma_{\tilde{T}}
                & \sigma_{\tilde{T}}^2
                & \rho_{\tilde{T}L}\,\sigma_{\tilde{T}}\,\sigma_{L} \\
            \rho_{\tilde{E}L}\,\sigma_{\tilde{E}}\,\sigma_{L}
                & \rho_{\tilde{T}L}\,\sigma_{\tilde{T}}\,\sigma_{L}
                & \sigma_{L}^2
          \end{bmatrix},
        \label{Eq 10: covariance matrix}
    \end{equation}
    where $\sigma_{\tilde{E}}$, $\sigma_{\tilde{T}}$, and $\sigma_{\log A}$ are the standard deviations of the three transformed quantities, and $\rho_{\tilde{E}\tilde{T}}$, $\rho_{\tilde{E}L}$, and $\rho_{\tilde{T}L}$ are the corresponding Pearson correlation coefficients \citep{Pearson1895} between $(\tilde{E},\tilde{T})$, $(\tilde{E},\log A)$, and $(\tilde{T},\log A)$, respectively (with $L$ denoting $\log A$).  
    In contrast to earlier work, we allow all three correlation coefficients $(\rho_{\tilde{E}\tilde{T}}, \rho_{\tilde{E}L}, \rho_{\tilde{T}L})$ to vary freely, so that the model can capture arbitrary Gaussian correlations between ellipticity, triaxiality, and size in the transformed space.
    
    We follow the same overall inference strategy as illustrated in Fig.~6 of \citet{Pandya2024}, but applied to our modified parametrization $(\vec{\mu},\Sigma)$ and to the ellipsoid library that is now constructed on a regular grid in $\vec{\gamma}$.  
    Because the logit transform maps $(0,1)$ to $(-\infty,\infty)$, we restrict the grid in $\logit E$ and $\logit T$ to the finite interval $[-7,7]$, which already corresponds to $E$ and $T$ values extremely close to 0 or 1 and thus comfortably encompasses all physically plausible shapes considered here.
    
    \subsubsection{Likelihood and prior specification}
    \label{section 2.3.3: PyMC setup and tests}
    
    Let $\vec{C} = \{C_k\}$ denote the observed bin counts in the $(q,\log a)$ histogram.  
    Given $(\vec{\mu},\Sigma)$, the model-predicted counts are $\lambda_k(\vec{\mu},\Sigma)$ as in Eq.~(\ref{Eq 7: model-hist}).  
    We adopt a Poisson likelihood for each bin,
    \begin{equation}
        \mathcal{L}(\vec{C} \mid \vec{\mu},\Sigma) 
        \;=\; \prod_k \mathrm{Poisson}\bigl(C_k \mid \lambda_k(\vec{\mu},\Sigma)\bigr),
        \label{Eq 11: Poisson likelihood}
    \end{equation}
    and aim to infer the posterior distribution
    \begin{equation}
        p(\vec{\mu},\Sigma\mid \vec{C}) \;\propto\; \mathcal{L}(\vec{C}\mid \vec{\mu},\Sigma)\, p(\vec{\mu},\Sigma),
        \label{Eq 12: posterior}
    \end{equation}
    where $p(\vec{\mu},\Sigma)$ encodes the prior information.
    
    We implement this model using the \texttt{PyMC} probabilistic-programming framework \citep{salvatier2015}.  
    This approach leverages automatic differentiation to compute exact gradients of the log-posterior density with respect to all nine free parameters $(\vec{\mu},\Sigma)$, enabling efficient exploration of the parameter space using Hamiltonian Monte Carlo (HMC), specifically the No-U-Turn Sampler (NUTS) algorithm \citep{Hoffman2014}.
    
    We adopt weakly informative priors for the model parameters.  
    For the mean parameters, we assign normal distributions,
    \begin{align}
        \mu_{\tilde{E}} &\sim \mathcal{N}(0,1^2), \\
        \mu_{\tilde{T}} &\sim \mathcal{N}(0,1^2), \\
        \mu_{\log A}    &\sim \mathcal{N}(0,0.5^2).
    \end{align}
    For the standard deviation parameters, we employ truncated normal priors.  
    This choice enforces positivity and prevents the sampler from exploring pathological regions where the intrinsic distribution becomes significantly narrower than the grid resolution of the pre-computed ellipsoid library.  
    Specifically, we define
    \begin{align}
        \sigma_{\tilde{E}} &\sim \mathcal{N}^+(1,0.9^2;0,7), \\
        \sigma_{\tilde{T}} &\sim \mathcal{N}^+(1,0.75^2;0.5,7), \\
        \sigma_{\log A}    &\sim \mathcal{N}^+(0,1^2;0,1),
    \end{align}
    where $\mathcal{N}^+(\mu,\sigma^2;a,b)$ denotes a normal distribution with mean $\mu$ and variance $\sigma^2$, truncated to the interval $[a,b]$.  
    The lower bound of $0.5$ for $\sigma_{\tilde{T}}$, for example, prevents the sampler from concentrating on triaxiality distributions that are too narrow to be resolved by the discrete library, while the upper bounds correspond to the limits of the grid.
    
    Our specific choice of prior parameters in the transformed $(\tilde{E}, \tilde{T})$ domain is guided by the behavior of the transformation itself.  
    A uniform prior in logit space would correspond to a prior in the physical $(E,T)$ space that is extremely sharply peaked near the boundaries 0 and 1.  
    We therefore adopt the above priors to approximate, in the transformed domain, the physical-space priors used by \citet{Pandya2024}, which were uniform in $(\mu_E,\mu_T,\mu_{\log A})$ and exponential in $(\sigma_E,\sigma_T,\sigma_{\log A}) \sim \exp(1)$.
    
    For the correlation structure, we adopt an LKJ prior \citep{Lewandowski2009} on the $3\times3$ correlation matrix $R$.  
    We decompose the covariance matrix as $\Sigma = D R D$, where $D$ is the diagonal matrix of standard deviations,
    \begin{equation}
        D = \mathrm{diag}(\sigma_{\tilde{E}}, \sigma_{\tilde{T}}, \sigma_{\log A}).
    \end{equation}
    The LKJ prior,
    \begin{equation}
        R \sim \mathrm{LKJ}(\eta),
    \end{equation}
    controls the concentration of $R$ around the identity matrix: $\eta = 1$ corresponds to a uniform prior over correlation matrices, while larger $\eta$ disfavors extreme correlations.  
    We adopt $\eta = 2$, which is still conservative but gently downweights $|\rho_{ij}| \simeq 1$.
    
    We run four independent NUTS chains on a GPU, each with 5000 tuning draws and 3000 posterior draws, and set \texttt{target\_accept=0.95} to allow the sampler to navigate the potentially complex posterior landscape.  
    Under these settings, the four chains converge to consistent posterior distributions according to standard diagnostics (effective sample size and $\hat{R}$), although several tenths of divergent transitions remain, indicating occasional exploration of regions near the boundary of positive definiteness for $\Sigma$ or other local pathologies in the geometry of the posterior.  
    We inspect the posterior samples and find no evidence that these divergences bias the inferred distributions.  
    In Appendix~\ref{Appendix A}, we present tests in which we apply the same inference framework to mock data drawn from known input parameters; NUTS succeeds in recovering most parameters accurately for sample sizes $N \gtrsim 1000$, while for smaller sample sizes, parameters involving triaxiality $(\mu_{\tilde{T} },\sigma_{\tilde{T}},\rho_{\tilde{E}\tilde{T}})$ are often only weakly constrained.  
    To improve the constraining power in some applications, we also consider variants of the model in which one or more of these triaxiality-related parameters are fixed to values measured directly from the simulation.
    
    In the rest of the paper, we refer to the original $(E,T,\log A)$-Gaussian model of \citet{Zhang2019} and \citet{Pandya2024} as the \textit{ETlogA} model, and to our transformed model as the \textit{ltETlogA} model.

    %%%%%%%%%%%%%%%%%%%%%%%%%%%%%%%%%%%%%%%%%%%%%%%%%%%%%%%%%%%%%%%%%
    % Figure for intrinsic 3D shape measurements
    \begin{figure*}
        \centering
    
        %================= Left column: centered test galaxy =================
        \begin{minipage}[c]{0.40\textwidth} % [c] = vertically centered
            \centering
            \caption*{(a) Test galaxy} % caption inside the panel (no subcaption counter)
            \includegraphics[width=\linewidth]{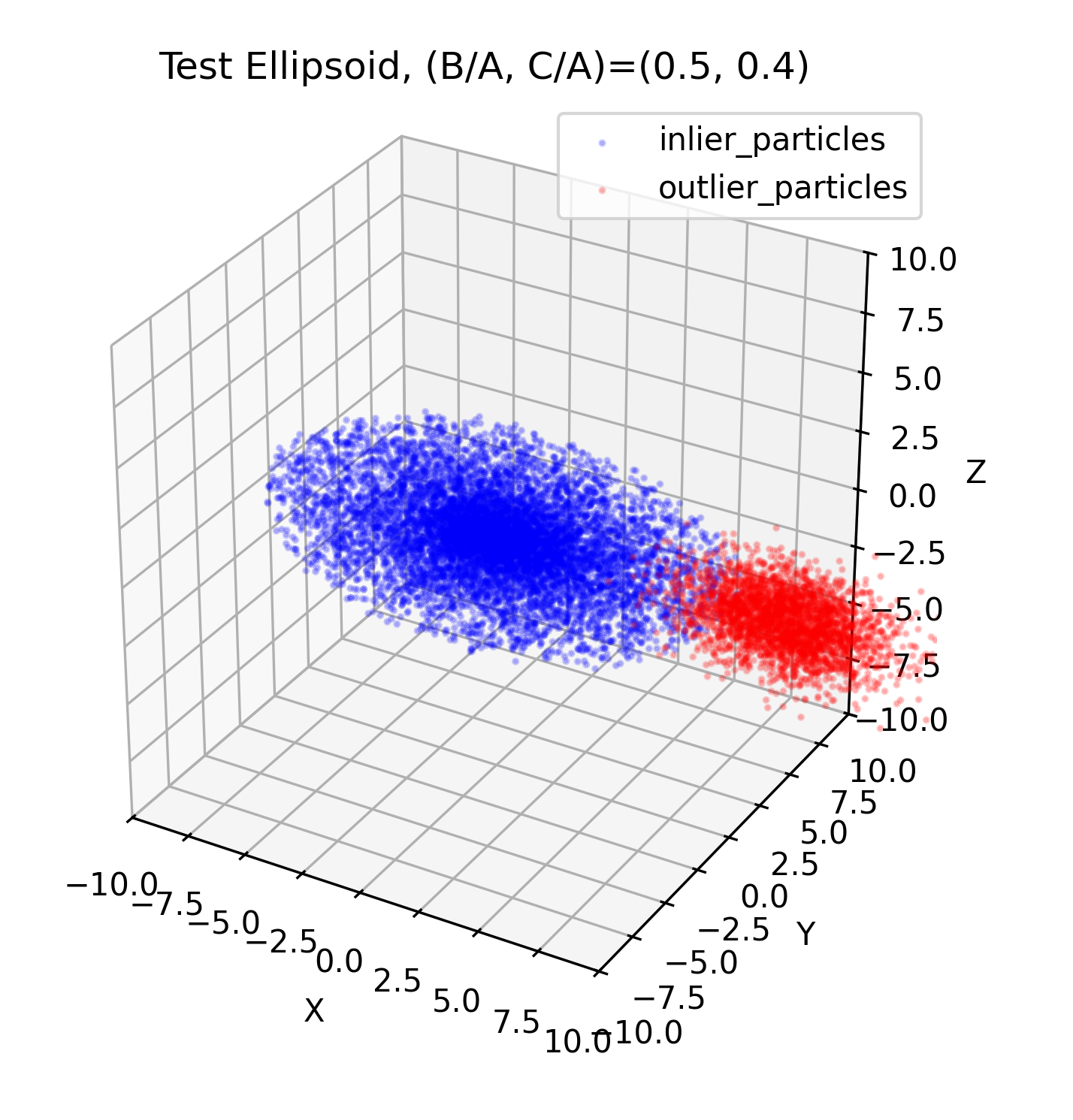}
        \end{minipage}%
        \hfill
        %================= Right column: two stacked recovery plots ==========
        \begin{minipage}[c]{0.55\textwidth}
            \centering
    
            % Top recovery plot
            \caption*{(b) Recovery without outliers}
            \includegraphics[width=\linewidth]{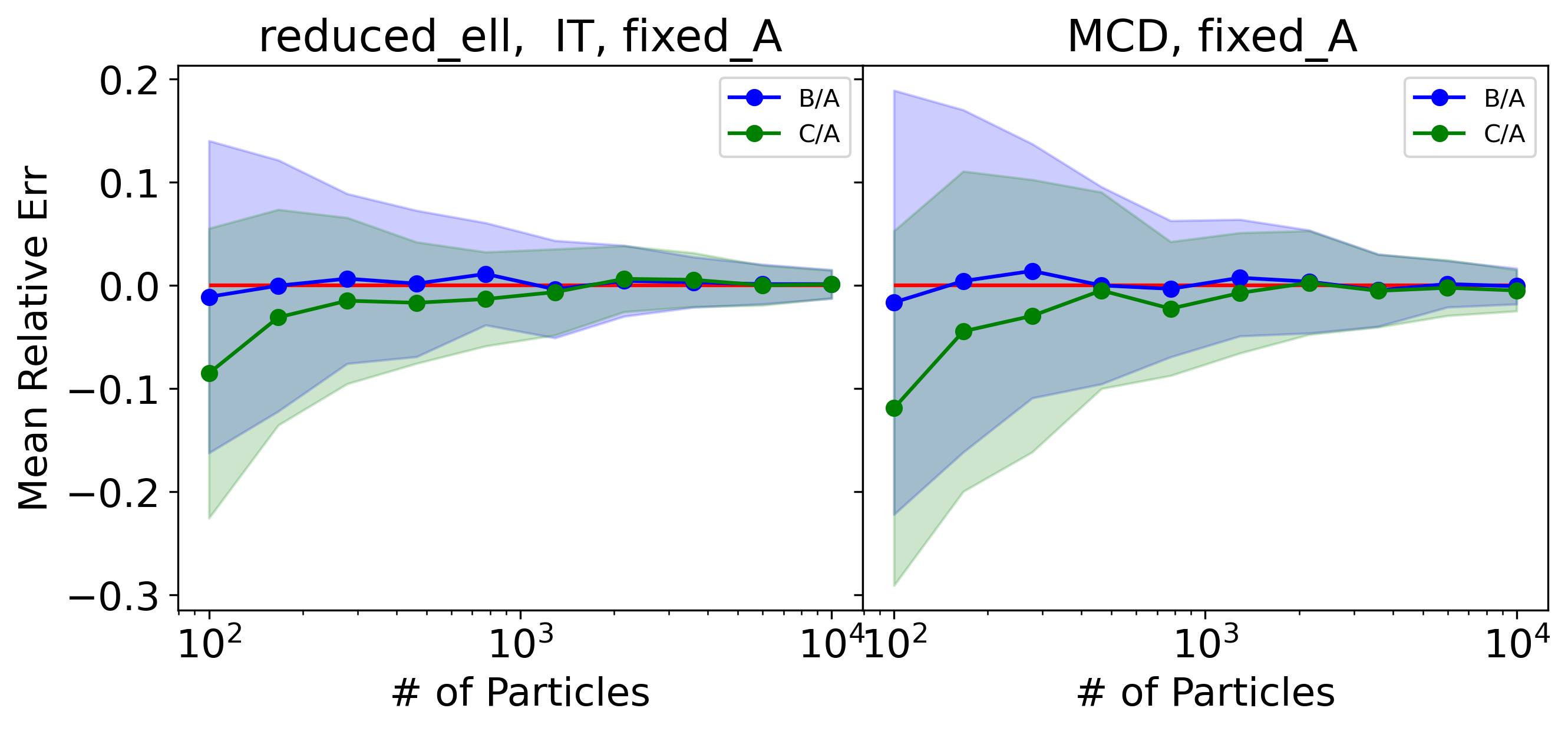}        
    
            % Tight vertical spacing between the two right panels
            \vspace{0.3\baselineskip}
    
            % Bottom recovery plot
            \caption*{(c) Recovery with outliers}
            \includegraphics[width=\linewidth]{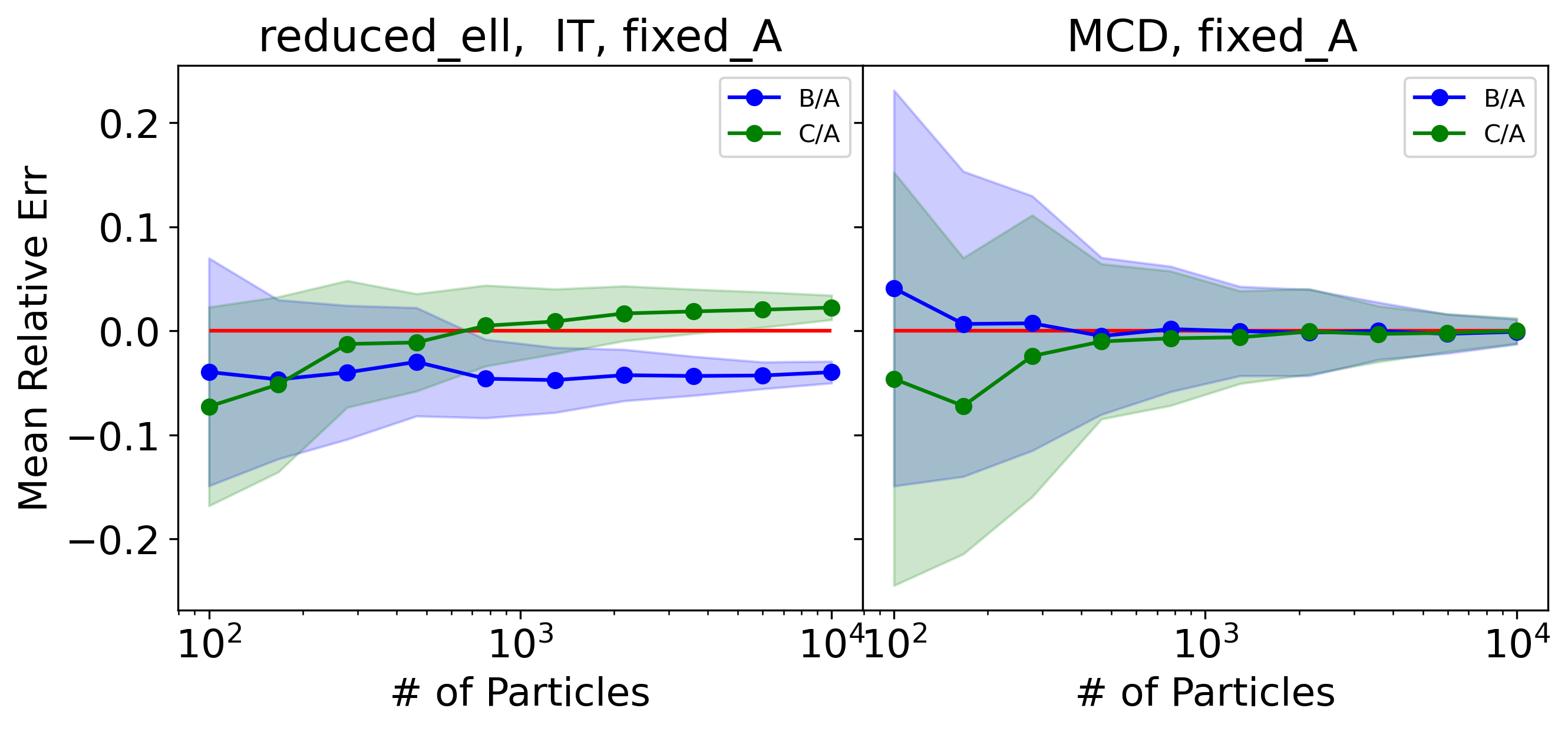}
    
        \end{minipage}
    
        \caption{
        Recovery tests for the intrinsic 3D shape of a test galaxy. 
        Panel (a): stellar distribution of the test ellipsoid, with inlier particles in blue and an added outlier component in red. 
        Panel (b): mean relative errors in the recovered axis ratios $B/A$ and $C/A$ as a function of particle number for the iterative reduced inertia–tensor (IT) method (left) and the robust MCD estimator (right), in the case without outliers; both methods are essentially unbiased for sufficiently large $N$, with the IT method showing slightly smaller scatter.
        Panel (c): same as (b) but with outliers included; the IT method is biased by the outliers, whereas the robust MCD estimator correctly recovers the intrinsic axis ratios of the main ellipsoid.
        }
        \label{IT/MCD test}
    \end{figure*}
    %%%%%%%%%%%%%%%%%%%%%%%%%%%%%%%%%%%%%%%%%%%%%%%%%%%%%%%%%%%%%%%%%%%%
    
\subsection{Intrinsic 3D Shape Measurements}
\label{section 2.4: Measuring 3D Shape}
    \subsubsection{Iterative inertia-tensor method}

    The intrinsic 3D shapes of simulated galaxies are often measured from the spatial distribution of their star particles using an iterative inertia-tensor approach.  
    This method (sometimes formulated as a “reduced inertia tensor’’) has been widely used to characterise halo and galaxy shapes in cosmological simulations (e.g. \citealt{Genel2015, Zhang2022, klein2026}; see also Sect.~2.2 of \citealt{Xia2026}).  
    In this subsection, we briefly summarise the standard procedure, which we use only as a reference and for comparison tests; our fiducial shape measurements are obtained with the robust estimator described in Sec.~\ref{section 2.4.2: MCD estimator} below.
    
    For a given galaxy, we first select star particles within a spherical aperture of radius $0.3\,R_{\mathrm{vir}}$ around the halo center, where $R_{\mathrm{vir}}$ is the virial radius. However, this preliminary selection presents two distinct challenges for shape estimation: the inclusion of distinct satellite galaxies within the aperture, and the fact that the stellar center of mass may not coincide with the dark matter halo center. We address these issues subsequently in Sec.~\ref{section 2.4.2: MCD estimator}.
    We then compute the reduced inertia tensor
    \begin{equation}
        M_{ij} = 
        \frac{\displaystyle\sum_n r_{\mathrm{ell},n}^{-2} \, m_n \, x_{n,i} x_{n,j}}
             {\displaystyle\sum_n m_n},
        \label{Eq 21: Reduced inertia tensor}
    \end{equation}
    where $m_n$ is the mass of the $n$-th star particle, $x_{n,i}$ is its $i$-th coordinate relative to the galaxy centre, and $r_{\mathrm{ell},n}$ is its elliptical radius,
    \begin{equation}
        r_{\mathrm{ell},n}^2 =
            x_{n,1}^2 +
            \frac{x_{n,2}^2}{(B/A)^2} +
            \frac{x_{n,3}^2}{(C/A)^2},
    \end{equation}
    evaluated using the axis ratios $(B/A,C/A)$. Diagonalising $M_{ij}$ yields three eigenvalues $\lambda_1 \geq \lambda_2 \geq \lambda_3$ and corresponding orthonormal eigenvectors.  
    The semi-axes of the best-fitting ellipsoid are proportional to the square roots of the eigenvalues,
    \begin{equation}
        A : B : C \;\propto\; \sqrt{\lambda_1} : \sqrt{\lambda_2} : \sqrt{\lambda_3},
    \end{equation}
    so that the intrinsic axis ratios are
    \begin{equation}
        \frac{B}{A} = \sqrt{\frac{\lambda_2}{\lambda_1}}, \qquad
        \frac{C}{A} = \sqrt{\frac{\lambda_3}{\lambda_1}}.
    \end{equation}
    
    We iteratively refine the shape measurements by recomputing $M_{ij}$ within an ellipsoidal volume whose axis ratios are given by $(B/A,C/A)$ from the previous iteration, keeping the semi-major axis fixed at $0.3\,R_{\mathrm{vir}}$ and rescaling the intermediate and minor axes accordingly.  
    The procedure is repeated until the axis ratios converge to within $10^{-4}$ between successive iterations.  
    In this way, we obtain intrinsic $(B/A, C/A)$ values and principal-axis directions for each galaxy.

    %%%%%%%%%%%%%%%%%%%%%%%%%%%%%%%%%%%%%%%%%%%%%%%%%%%%%%%%%%%%%%%%%%%%
    \subsubsection{Robust minimum covariance determinant estimator}
    \label{section 2.4.2: MCD estimator}
    
    As expected, the iterative inertia–tensor (IT) method works well once obvious subhaloes and other outliers have been removed from the sample ( \citealt{Lucas2024}).  
    However, its sensitivity to outliers motivates a more robust alternative.  
    We note that the mass/inertia tensor in Eq.~(\ref{Eq 21: Reduced inertia tensor}) is proportional to a (weighted) sample covariance matrix, whose eigenvectors and eigenvalues define the principal axes and axis lengths of the best-fitting ellipsoid.
    
    We therefore also iteratively apply a robust minimum covariance-determinant (MCD) estimator \citep{Rousseeuw1984} to the 3D positions of star particles.  
    The MCD estimator searches for a subset of the data whose covariance matrix has the smallest determinant and uses this subset to compute robust estimates of the centroid $\hat{\boldsymbol{\mu}}$ and covariance matrix $\hat{\Sigma}$.  
    We convert the eigenvectors and eigenvalues of $\hat{\Sigma}$ into $(A,B,C)$ and $(B/A,C/A)$ in exactly the same way as for the standard inertia-tensor method.  
    This procedure downweights or excludes spatial outliers (e.g. substructures or tidal features) and returns shapes characteristic of the main stellar body.
    
    To quantify the performance of the IT and MCD estimators, we perform controlled tests with mock galaxies, both with and without outliers.  
    We construct a “test galaxy’’ centered at the origin whose stellar density follows an isothermal profile $\rho(r_{\mathrm{ell}}) \propto r_{\mathrm{ell}}^{-2}$ with unit particle masses, truncated at $r_{\mathrm{ell}} = 10$.  
    The intrinsic axis ratios are set to $(B/A,C/A) = (0.5,0.4)$.  
    We then add an outlier population whose particles are drawn from a trivariate Gaussian centered at $(x,y,z) = (12,0,0)$ with intrinsic axis ratios $(B/A,C/A) = (0.5,0.4)$ (Fig.~\ref{IT/MCD test}a).  
    The absolute spatial scale is irrelevant here because we are only interested in the recovery of the axis ratios.
    
    To assess bias and scatter, we generate 100 independent realizations of the stellar distribution for each configuration (different random seeds) and measure the axis ratios with both estimators, keeping the major axis fixed at $A = 15$, which encloses the majority of the particles, analogous to our choice of $0.3\,R_\mathrm{vir}$ for the actual galaxies.  
    We repeat this test for galaxies containing between $N = 100$ and $N = 10^4$ star particles after the initial spherical aperture cut.  
    Figure~\ref{IT/MCD test} (panels b and c) shows the mean relative errors and their scatter as a function of $N$, where for each axis ratio we define
    \begin{equation}
        \mathrm{Relative~Error}
        = \frac{(B/A)_{\mathrm{est}} - (B/A)_{\mathrm{true}}}{(B/A)_{\mathrm{true}}}
    \end{equation}
    (and analogously for $C/A$).

    %%%%%%%%%%%%%%%%%%%%%%%%%%%%%%%%%%%%%%%%%%%%%%%%%%%%%%%%%%%%%%%%%%%%%%%%%%%%%%%%%%%%%%%%%%%%%
    % Mass-Size Relation
    \begin{figure*}[!ht]
        \centering
        \includegraphics[width=1\linewidth]{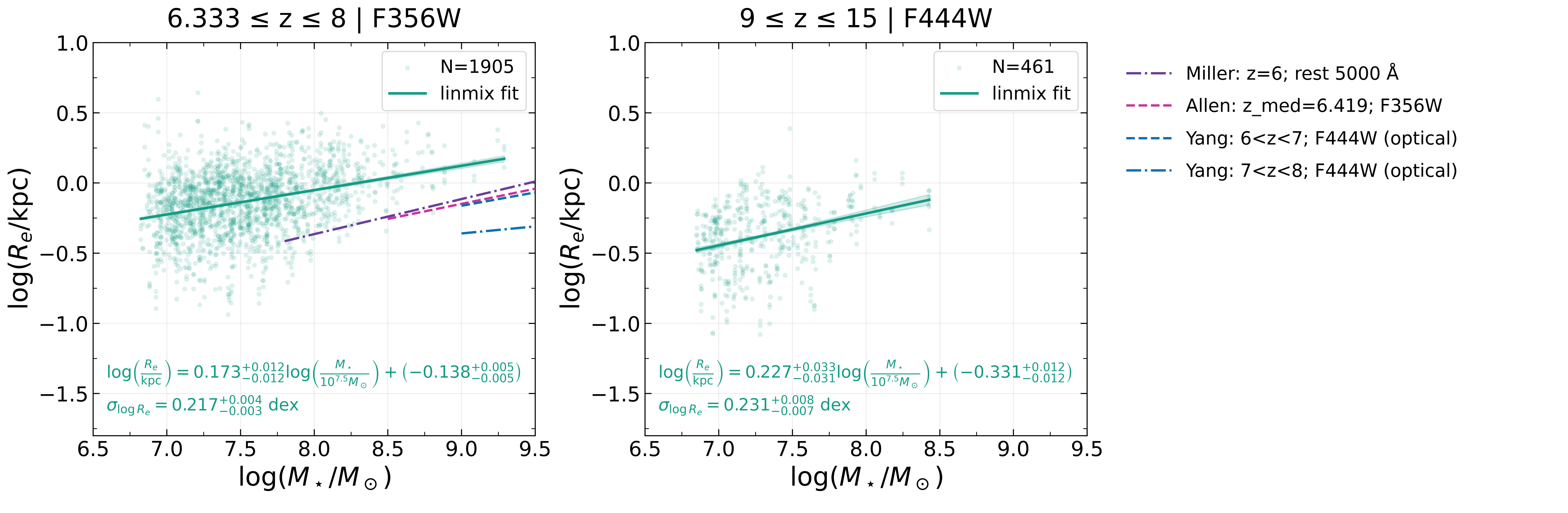}
        \caption{The rest-frame optical mass-size distribution of our galaxies. The green solid lines indicate the median mass-size relation, the shaded regions are the 68th\% credible intervals of the fit. 
        We provide the fitted mass-size relation with pivot mass $\Mstar=7.5$ and its scatter at the bottom of the plots.
        Observational results in similar redshift bins are overplotted in dark violet \citep{Miller2025}, magenta \citep{Allen2025}, and blue \citep{Yang2025}.
        }
        \label{fig: mass-size relation}
    \end{figure*}
    
    In the case without outliers, both methods are essentially unbiased once the initial number of particles $N$ exceeds $\sim 10^3$, with the IT method exhibiting slightly smaller scatter than the robust MCD estimator (Fig. \ref{IT/MCD test}b).  
    When outliers are added, however, the IT results are clearly driven by the outlier component, whereas the robust MCD estimator correctly recovers the shape of the main ellipsoid (Fig. \ref{IT/MCD test}c).  
    From these tests we also infer that, for $N \approx 10^3$, the $1\sigma$ scatter around the mean relative error is typically $\sim 5\%$.  
    To avoid retaining galaxies for which the MCD estimator discards a large fraction of the stellar distribution or leaves too few particles for a reliable shape measurement, we additionally require that the fraction of star particles retained after the final MCD iteration satisfies
    \begin{equation}
        \frac{N_{\mathrm{end}}}{N} \geq 0.5
        \quad\text{and}\quad
        N_{\mathrm{end}} \geq 1000,
        \label{Eq 26: 3D shapes cut}
    \end{equation}
    where $N_{\mathrm{end}}$ is the number of particles used in the final MCD estimate and $N$ is the number of particles within the initial spherical aperture.  
    These cuts are applied when we use the robust MCD estimator on the actual simulation data.
    
    Although fixing the major-axis length is our default choice, we have also experimented with fixing the ellipsoidal volume instead, and with alternative weighting schemes in Eq.~(\ref{Eq 21: Reduced inertia tensor}) (e.g. different powers of $r_{\mathrm{ell},n}$).  
    The adopted configuration yields the smallest bias for the test galaxy considered here, and the qualitative trends are consistent with the behavior reported by \citet{Lucas2024}.

%%%%%%%%%%%%%%%%%%%%%%%%%%%%%%%%%%%%%%%%%%%%%%%%%%%%%%%%%%%%%%%%%%%
% 2D shape results
\begin{figure*}
    \centering
    \includegraphics[width=\textwidth]{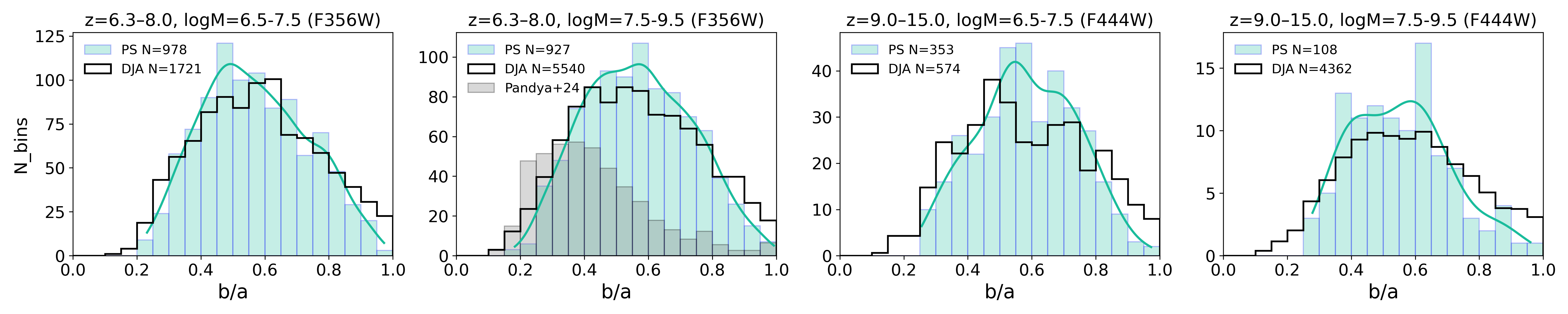}
    \includegraphics[width=\textwidth]{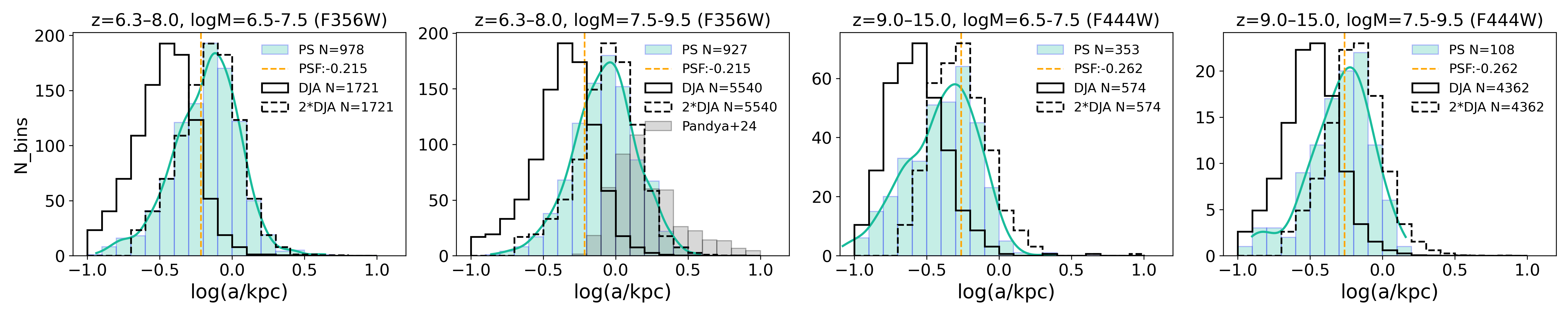}
    \includegraphics[width=\textwidth]{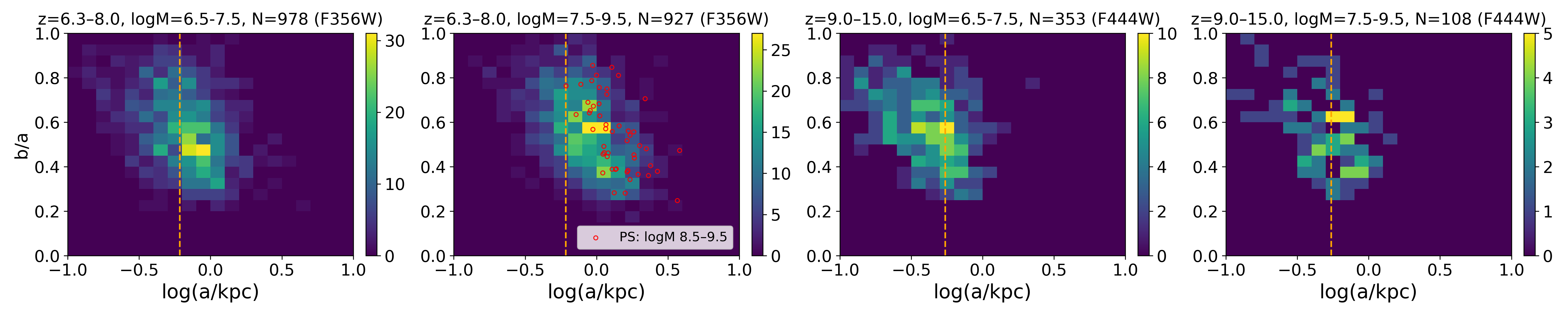}
    \caption{
    Projected 2D structural properties of \FIRE galaxies compared to galaxies in the Dawn JWST Archive (DJA) catalog.
    Columns correspond to the four stellar-mass--redshift bins in Table~\ref{Final Data}, ordered from low to high redshift.
    \textbf{Top row:} marginal distributions of projected axis ratio $b/a$ for \FIRE galaxies (filled histograms) and DJA galaxies (black step histograms), renormalized to the same total counts within each bin.
    \textbf{Second row:} marginal distributions of projected major-axis size $\log(a/\mathrm{kpc})$ for the same samples; dashed black steps show the DJA size distributions after multiplying $a$ by a factor of two (i.e. shifting $\log a$ by $+0.3$\,dex) to facilitate the comparison.
    The vertical orange dashed line indicates the effective PSF FWHM in the corresponding band.
    Grey histograms (where present) show the $z=3$--8, $\log(M_\star/\Msun)=9$--9.5 dwarf galaxy bin from \citet{Pandya2024} ($z_{\rm med}\simeq 3.5$).
    \textbf{Third row:} joint distributions of $(b/a,\log a)$ for the \FIRE sample. The red circles are galaxies with stellar mass $\log(M_\star/\Msun)=8.5$--9.5 and there are about 50, illustrating most of our galaxies are in the lower mass region. 
    }
    \label{2d shape properties}
\end{figure*}

% \clearpage
\begin{figure*}
    \centering
    \includegraphics[width=\textwidth]{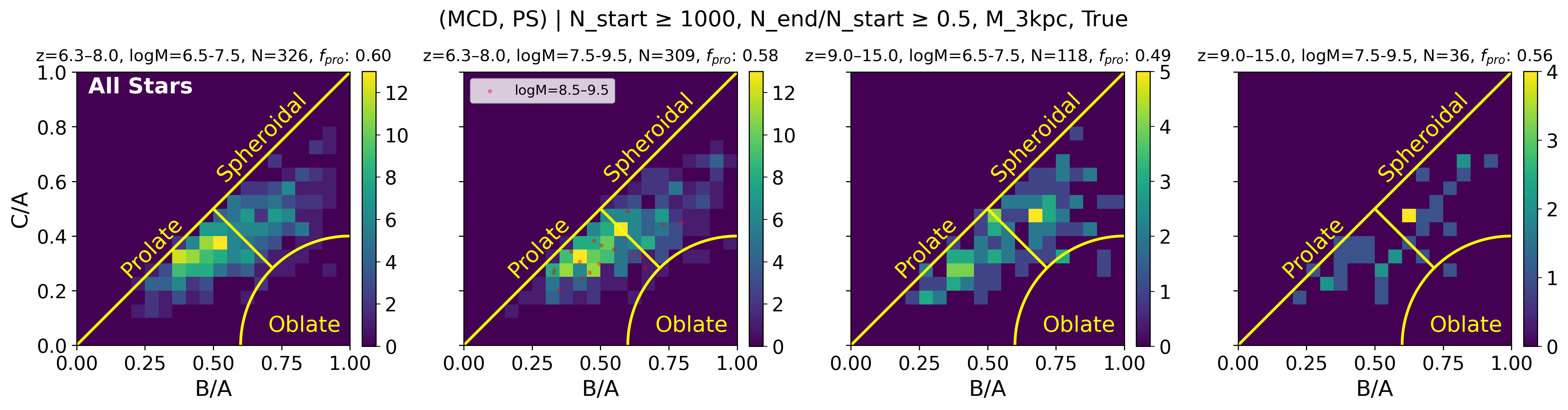}
    \includegraphics[width=\textwidth]{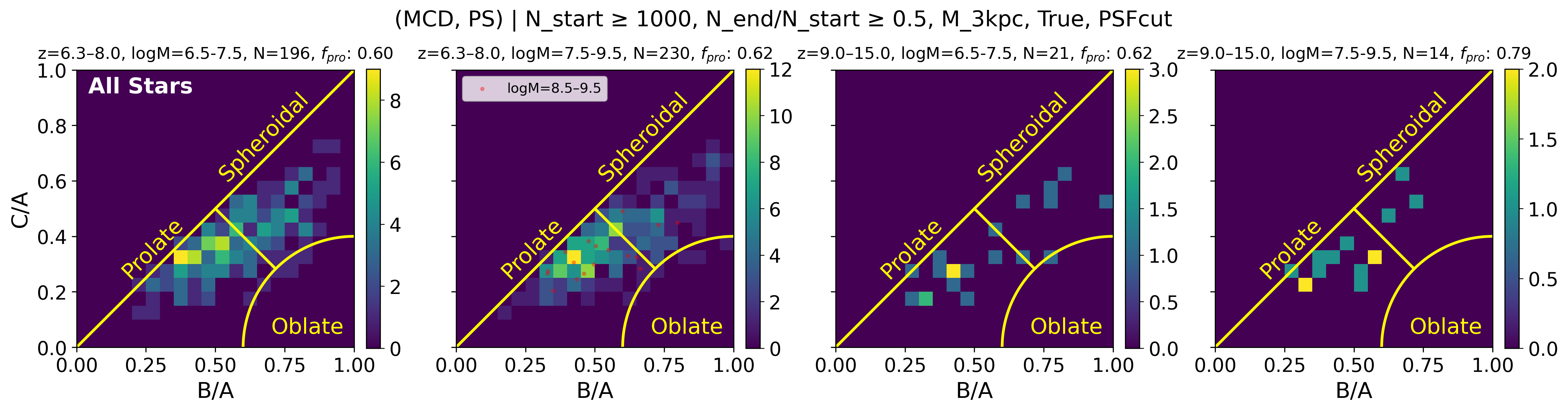}
    
    \caption{
    Intrinsic axis-ratio distributions measured with the robust MCD estimator.
    Panels show 2D histograms in the $(C/A,\,B/A)$ plane for the four stellar mass--redshift bins in Table~\ref{Final Data}; the yellow boundaries indicate the oblate/prolate/spheroidal regions following the classification of \citet{Zhang2019}. The prolate fraction in each bin, $f_{\rm pro}$, is reported in the panel titles.
    \textbf{Top row:} results for all galaxies passing the MCD quality cuts.
    \textbf{Bottom row:} the same, but restricted to galaxies that are resolved in projection according to the PSF--FWHM size criterion (Sect.~\ref{section 2.2: Mock Observation}).
    The intrinsic distributions are broadly similar with and without the PSF-based selection, although the resolved samples at the highest redshifts are small.
    }
    \label{measured 3D shapes}
\end{figure*}

\begin{figure*}
    \centering
    \includegraphics[width=\textwidth]{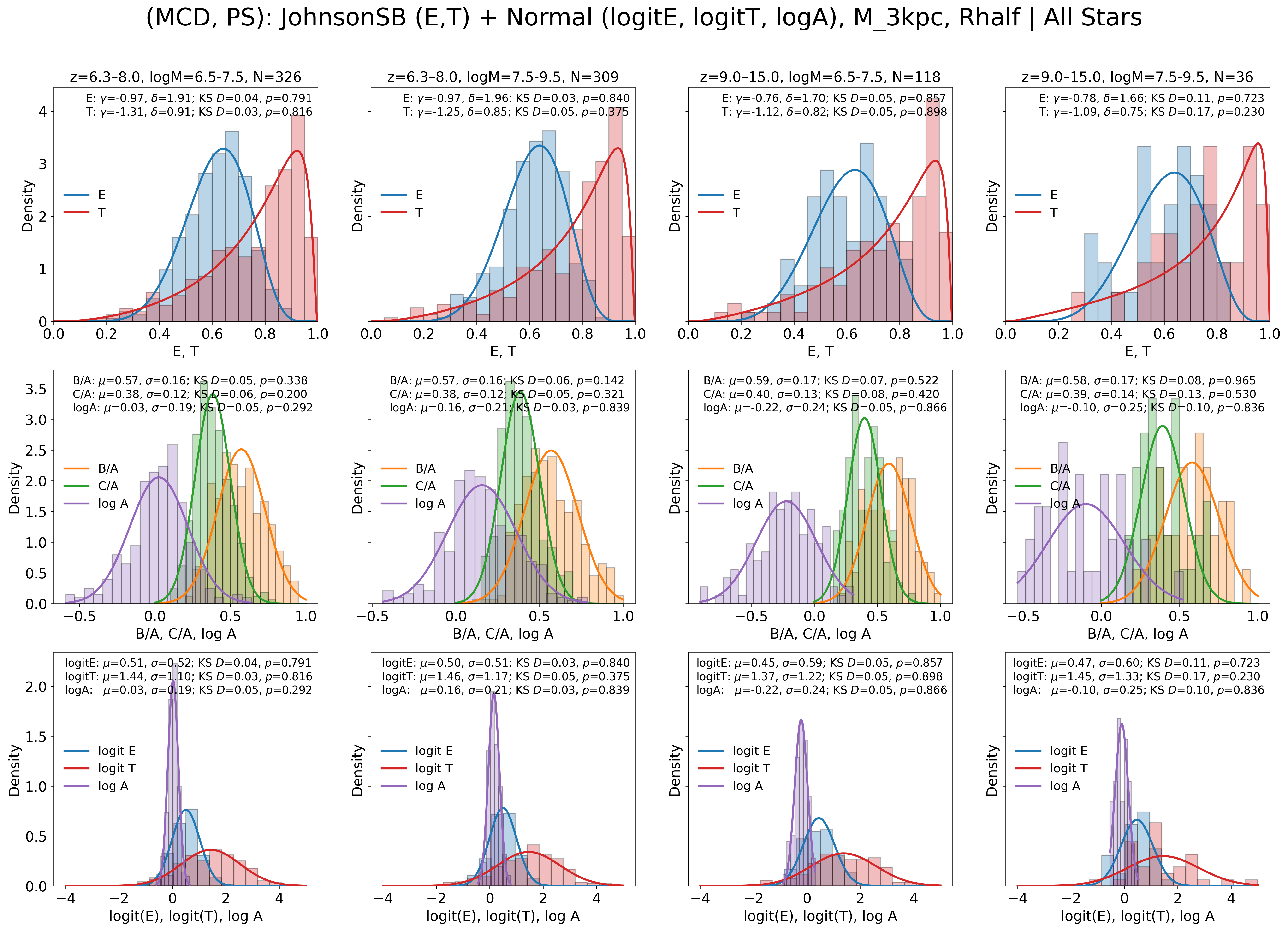}
    \caption{
    Marginal distributions of intrinsic 3D shape parameters in the four mass--redshift bins, measured with the robust MCD estimator and overlaid with best-fitting analytical models.  
    Inset boxes report the best-fit parameters and the one-sample Kolmogorov--Smirnov (KS) goodness-of-fit statistics ($D$ and $p$-value).
    \textbf{Top row:} Distributions of intrinsic ellipticity ($E$) and triaxiality ($T$), each fitted with a Johnson~SB distribution.  
    The small $D$ values and large $p$-values indicate that Johnson~SB provides an excellent description of the simulated populations, with only weak variation across bins.
    \textbf{Middle row:} Distributions of the physical axis ratios $(C/A,\,B/A)$ and intrinsic size $\log A$, fitted with Gaussian models.  
    While these fits capture the overall shapes of the marginals, the typically lower $p$-values may suggest a statistically poorer description than the Johnson~SB fits for $(E,T)$.
    \textbf{Bottom row:} Distributions of the transformed variables $(\tilde{E},\,\tilde{T},\,\log A)$, where $\tilde{E}\equiv\logit(E)$ and $\tilde{T}\equiv\logit(T)$, fitted with Gaussians.  
    Because a Johnson~SB variate is equivalent to a logit-normal variate by construction, the goodness-of-fit metrics mirror those in the top row. This behavior motivates the Gaussian assumption in the transformed space adopted by our $\ltETa$ inference model.}
    \label{marginal fitted distribution}
\end{figure*}

%%%%%%%%%%%%%%%%%%%%%%%%%%%%%%%%%%%%%%%%%%%%%%%%%%%%%%%%%%%%%%
\section{Results}
\label{section 3: Results}
    In this section we present the projected 2D and intrinsic 3D morphologies of the simulated \FIRE galaxies.
    Unless stated otherwise, we restrict the analysis to central galaxies and require that each system has reliable measurements from both the 2D S\'ersic fitting and the 3D shape analysis, so that the same objects are used for all comparisons.
    
    We subdivide the sample into two bins in stellar mass and redshift, with final sample sizes listed in Table~\ref{Final Data}.
    Throughout this paper, stellar masses are defined as the total mass in star particles within a spherical aperture of radius $3\,\mathrm{kpc}$ (proper). The aperture size is motivated by JWST observations which show that two-dimensional aperture diameters of 0.3 and 0.7 arcsec (1.8 and 4 kpc at z = 6) typically contain 70–80 per cent and 85--90 per cent of the total flux \citep{Naidu2022, Finkelstein2023, Tacchella2023, Donnan2022, Adams2024}.
    For each bin we apply the quality cuts on 3D shape measurements described in Sect.~\ref{section 2.4.2: MCD estimator}, namely
    $N_\mathrm{end} > 10^3$ and $N_{\mathrm{end}}/N \geq 0.5$,  
    where $N_{\mathrm{end}}$ is the number of particles used in the final MCD estimate and $N$ is the number of particles within the initial spherical aperture.
    Where indicated, we further impose a cut on projected size relative to the PSF FWHM.
    
    \begin{table}[!ht]
        \caption{Final sample sizes by mass–redshift bin.}
        \label{Final Data}
        \centering  
        \begin{tabular}{c c c c c}
        \hline\hline
        $z$     & $\log (M_\star/\Msun)$ & Filter & PySersic & MCD \\
        \hline
        6.3--8  & 6.5--7.5               & F356W  & 978 (592) & 326 (196) \\
        6.3--8  & 7.5--9.5               & F356W  & 927 (682) & 309 (230) \\
        9--15   & 6.5--7.5               & F444W  & 353 (80)  & 118 (21)  \\
        9--15   & 7.5--9.5               & F444W  & 108 (41)  & 36  (14)  \\
        \hline
        \end{tabular}
        \vspace{0.2cm}
    
        \tablefoot{
          \small
          Columns list redshift range, stellar-mass range, NIRCam filter, and the number of galaxies with successful 2D S\'ersic fits (PySersic) and robust 3D shape measurements (MCD), after all quality cuts.  
          Numbers in parentheses indicate the subset that additionally satisfy the PSF–FWHM size cut used in some parts of the analysis.
        }
    \end{table}

    \subsection{Mass-Size relation}
    \label{mass-size relation}

    We first start with an impression of where our samples are located on the mass-size axes with a mass-size relation overplotted as shown in Figure \ref{fig: mass-size relation}.
    In each mass-redshift bin, we fit a linear relation between the stellar mass ($M_{\star}$) and half-light radius ($R_e$) in log space
    \begin{equation}
        \log\frac{R_e}{kpc}=\alpha + \beta \log\frac{M_\star}{10^{7.5}M_\odot}+\epsilon
        \label{eq: mass-size relation}
    \end{equation}
    where $\alpha$ is the y-intercept at the pivot stellar mass $\Mstar=7.5$ and $\beta$ is the slope.
    $\epsilon$ is the intrinsic scatter in log half-light radius at each given stellar mass obeying a log-normal distribution $\mathcal{N}(0, \sigma_{\log R_e}^2)$.
    The fit is done using the Python package \texttt{Linmix} \citep{Kelly2007}, a hierarchical Bayesian model for fitting a straight line.

    The median redshifts of the lower- and higher-redshift samples are \(z_{\rm med}=7\) and \(9\), respectively. As shown in Figure \ref{fig: mass-size relation}, the \FIRE galaxies follow a size–mass relation with a slope broadly comparable to those reported for observed high-redshift star-forming galaxies \citep{Miller2025, Allen2025}. 
    At fixed stellar mass, however, their projected half-light radii are systematically larger by approximately \(0.3\) dex.
    Several effects may contribute to this offset.
    First, the simulated galaxies may have intrinsically a more extended distribution of stellar light compared with observed galaxies.
    Secondly, the plotted relations may differ because of differences in sample selection and systematics in the inferred stellar masses.
    Finally, our mock images do not account for noise and various systematics that may be present in realistic images.
    
    % An earlier analysis of FIREbox found that low-mass galaxies have stellar half-mass radii approximately \(0.3\) dex larger than observed effective radii at \(z=0\) \citep{Feldmann2023}, although this comparison uses different size definitions and cannot be directly extrapolated to the present redshift range.
    % A matched comparison that applies the observational selection and measurement procedures to the mock images is needed to determine the relative contributions of these effects. 
    
    Because our primary goal is to infer the shapes of low-mass galaxies from the joint distribution of projected axis ratios and sizes, the systematic offset in the absolute size scale is unlikely to affect our main conclusions substantially.

    \subsection{Projected 2D shape properties}
    \label{2D shapes}
    
    Figure~\ref{2d shape properties} summarizes the projected 2D structural properties of the \FIRE galaxies and compares them to those of galaxies in the Dawn JWST Archive (DJA) catalog \citep{Genin2025}.  
    The DJA catalog combines sources from several deep JWST surveys; photometric redshifts are derived with \texttt{EAZY} \citep{Brammer2008}, and band-invariant S\'ersic parameters $(R_e,b/a,n)$ are measured with \texttt{SourceXtractor++} (SE++; \citealt{Bertin2020, Kummel2022}).
    
    In the first two rows, the filled histograms show the marginal distributions of projected axis ratio $b/a$ and projected major-axis size $\log(a/\mathrm{kpc})$ from the \texttt{PySersic} fits to the mock \FIRE images.  
    The black step histograms show the corresponding DJA distributions, renormalized to the same total counts within each mass--redshift bin. 
    The shape distribution of simulated galaxies is in approximate agreement with those in the DJA catalog but slightly more strongly peaked. 
    The dashed black steps in the second row show the DJA size distributions after multiplying $a$ by a factor of two (i.e. shifting $\log a$ by $+0.3$\,dex), highlighting that the sizes of simulated galaxies are on average larger than those in the DJA catalog but with an otherwise similar distribution.
    The vertical orange dashed line marks the effective PSF FWHM in the corresponding band and therefore indicates the approximate resolution scale for size measurements.  
    Grey histograms (where present) show the $z=3$ -- 8, $\Mstar=9$ -- 9.5 dwarf galaxy samples from \citet{Pandya2024}, whose median redshift ($z_{\rm med}\simeq 3.5$) is substantially lower than that of our simulated bins.
    
    The third row shows the joint distributions in the $(b/a,\log a)$ plane for \FIRE.  
    We see a clear negative correlation between $b/a$ and $\log{a}$, which suggests a substantial contribution from prolate systems. 
    
    % It should be noted that we are comparing the noise-free distributions of \FIRE with the noise-present distributions recorded in the DJA catalog. The median errors of the DJA catalog sources are $\Delta(b/a)\sim0.02$ and $\Delta(\log{a})\sim0.1$, which makes only a small difference due to our small histogram bin number.

    \subsection{Measured 3D shapes}
    \label{section 3.2: 3D intrinsic shapes}
    In this section, we characterize the intrinsic 3D morphologies of simulated \FIRE galaxies by employing the robust MCD estimator (Sect.~\ref{section 2.4.2: MCD estimator}) and use these measurements to assess the assumptions underlying the forward-modeling Bayesian inference framework.

    Figure~\ref{measured 3D shapes} shows the distributions of intrinsic axis ratios in the $(C/A,\,B/A)$ plane. The top row includes all galaxies that pass the MCD quality cuts; the annotated quantity $f_{\rm pro}$ denotes the fraction of galaxies classified as prolate in each mass--redshift bin, following the partitioning scheme of \citet{Zhang2019}. The distributions indicate that, in the lower-redshift bins, \FIRE dwarf galaxies are predominantly prolate ($f_{\rm pro}\sim{}0.6$) with a substantial spheroidal component and very few clearly oblate systems, consistent with trends reported in other simulation suites (e.g. VELA, TNG50, FIREbox). Toward higher redshift, the prolate and spheroidal components become more comparable in relative importance. We caution that these comparisons are made across bins rather than for tracked progenitors, and therefore do not by themselves establish evolutionary pathways.
    
    The second row shows the same distributions after excluding galaxies that are unresolved in projection according to the PSF--FWHM size criterion (Sect.~\ref{section 2.2: Mock Observation}). The intrinsic-shape distributions are broadly similar to those of the full sample, suggesting that the MCD-based shape measurements are only weakly sensitive to this resolution selection. However, the PSF-based cut can substantially modify the projected $(b/a,\log a)$ distributions (Sect.~\ref{2D shapes}), and is therefore expected to impact the inferred intrinsic distributions in a forward-modeling analysis. Owing to the small number of resolved galaxies in the two highest-redshift bins, we do not attempt to interpret the PSF-selected samples in those cases.

    We define the intrinsic size, $A$, as the elliptical half-mass radius--specifically, the semi-major axis of the ellipsoid defined by the axis ratios $B/A$ and $C/A$ (determined via the robust MCD estimator) that encloses half of the total mass inside. With these measured intrinsic shapes in hand, we can directly evaluate four common modeling assumptions in the Bayesian framework
    
    First, we test whether the intrinsic parameters for our samples are well described by a multivariate normal distribution in the original $(E, T, \log A)$ space. Figure~\ref{marginal fitted distribution} compares the measured marginal distributions under several parameterizations and quantifies the goodness of fit using one-sample Kolmogorov--Smirnov tests \citep{Kolmogorov1933, Smirnov1948}. We find that the bounded parameters $(E,T)$ are better captured by a JohnsonSB distribution \citep{Johnson1949} than by a Gaussian. This motivates working with the transformed variables
    \begin{equation}
        \tilde{E} \equiv \logit(E), \qquad \tilde{T} \equiv \logit(T),
    \end{equation}
     where $\logit{x}$ = $\ln{\frac{x}{1-x}}$, for which the corresponding marginals are close to normal, providing the basis for our \textit{ltETlogA} model. An alternative approximation is to assume that $(B/A,\,C/A,\,\log A)$ follow a trivariate Gaussian; while this is less accurate than the \textit{ltETlogA} description for our samples, it can still be a useful practical approximation given the bounded nature of $(C/A,\,B/A)$.
    
    Second, we assess whether a single population model is adequate within each bin. For the dwarf galaxy samples considered here, the measured marginals do not show strong evidence for distinct multi-component structure, and a single-population description is sufficient at our current sample sizes. We note, however, that this need not hold for other datasets or selection functions.
    
    Third, we revisit the correlation assumptions. Because $(C/A,\,B/A)$ map deterministically to $(E,T)$, correlations in axis-ratio space imply correlations in $(E,T)$ (and vice versa). In our samples, the Pearson correlations among $(\tilde{E},\,\tilde{T},\,\log A)$ are all positive, motivating a model that allows all three correlation coefficients
    \(\rho_{\tilde{E}\tilde{T}}\), \(\rho_{\tilde{E}L}\), and \(\rho_{\tilde{T}L}\) (with $L\equiv \log A$) to vary freely, rather than restricting the covariance structure to a single non-zero $\rho_{EL}$.

    Finally, we show that through Binney's projection formalism, we deduce unbiased projected axis-ratios $b/a$ compared to S\'ersic measured axis-ratios as shown in Appendix \ref{Appendix E}. 
    The half-mass radius we measured in simulation is however, slightly biased compared to S\'ersic fitted half-light radius, as expected.
    The inconsistency between two definitions of the stellar sizes suggests using a light-weighted measurement.
    In the next section, we will see that the intrinsic shape distribution can be reasonably recovered even with the presence of bias, as long as enough samples are provided ($N\gtrsim1000$).  

    \begin{figure*}[!ht]
        \centering
        \includegraphics[width=\textwidth]{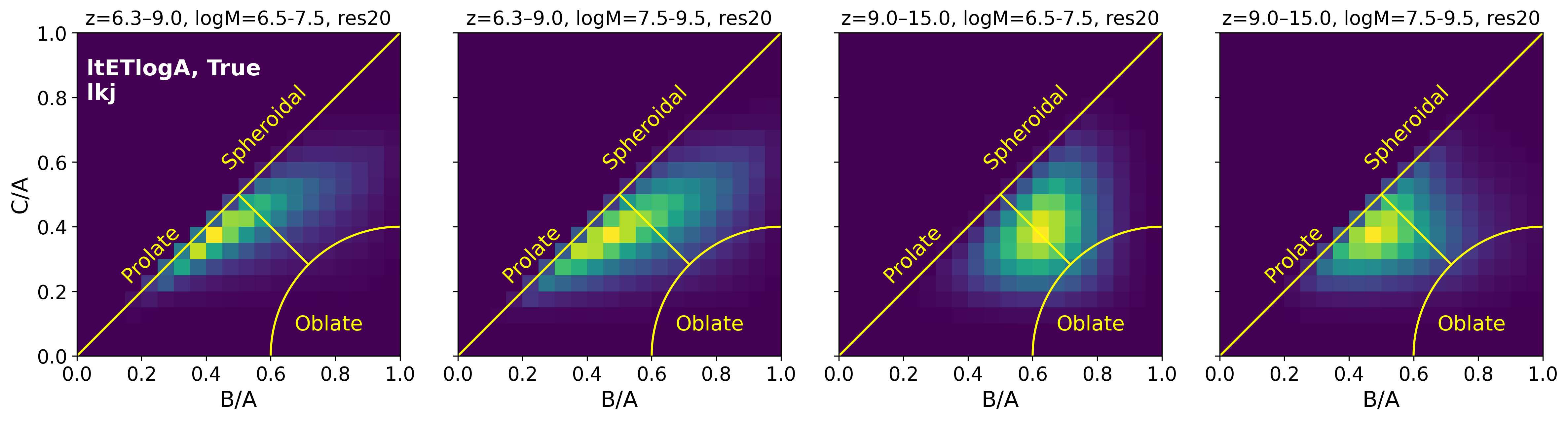}
        \includegraphics[width=\textwidth]{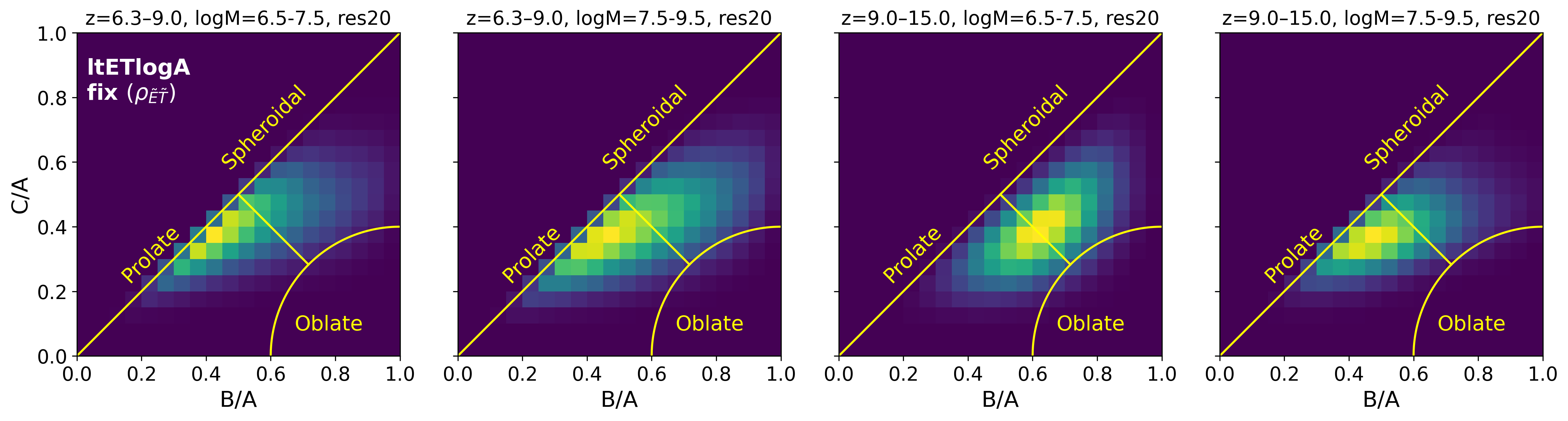}
        \includegraphics[width=\textwidth]{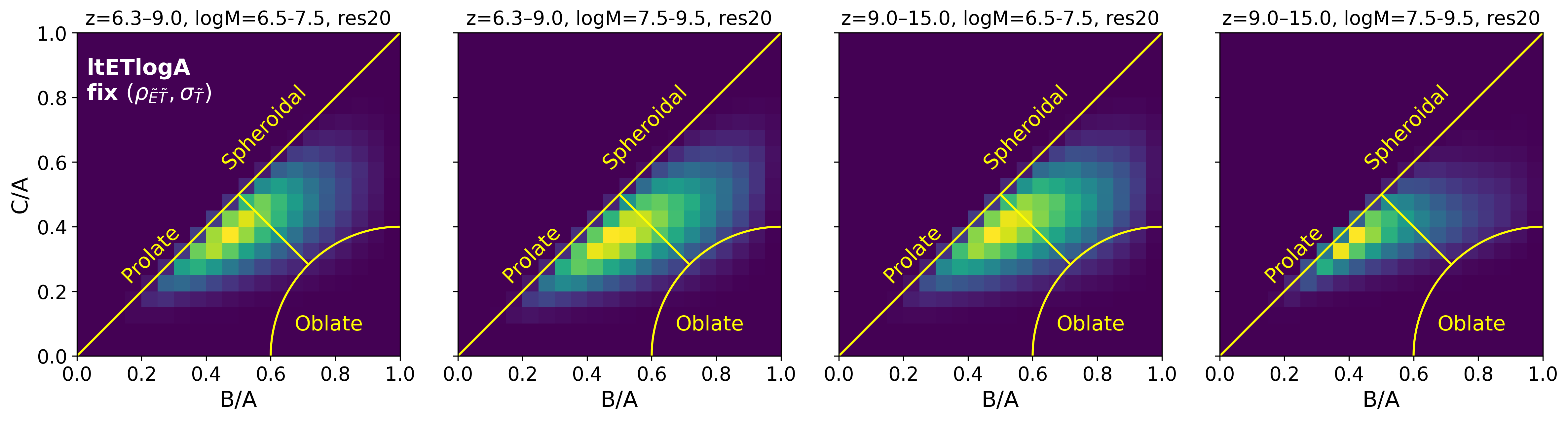}
        \includegraphics[width=\textwidth]{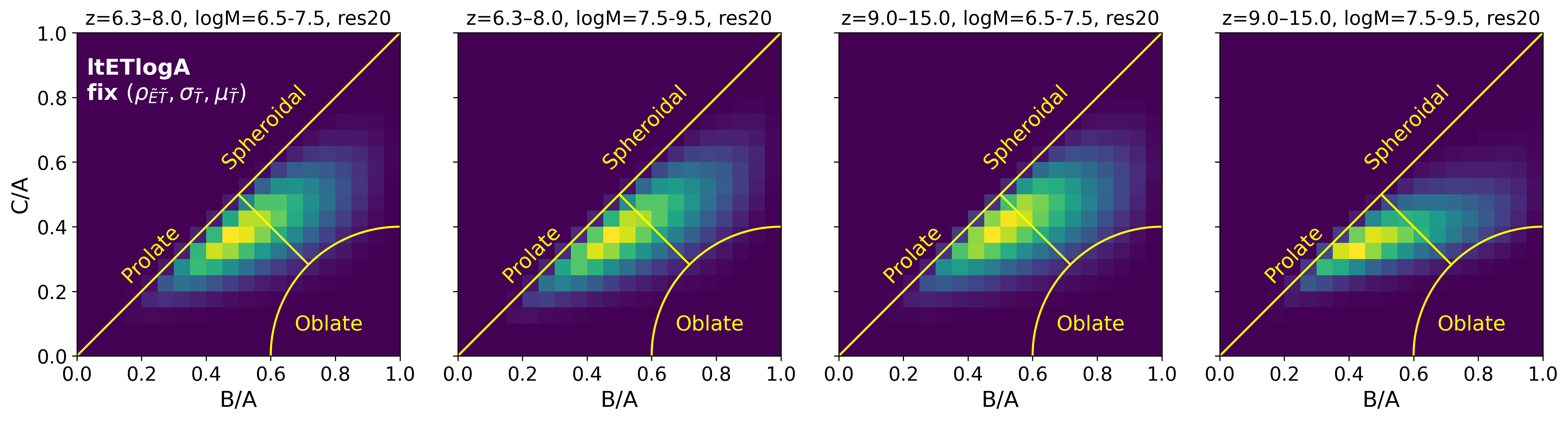}
    
        \caption{Inferred intrinsic 3D shape distributions in the $(C/A,\,B/A)$ plane for the four stellar-mass--redshift bins (columns), using the $\ltETa$ model. 
        Color indicates the model-predicted count; the solid lines mark the boundaries of the oblate, prolate, and spheroidal regions following the classification scheme of \citet{Zhang2019}. 
        Rows illustrate the impact of progressively fixing triaxiality-related parameters to the particle-measured (“true”) values.
        \textbf{Row 1:} all parameters are free.
        \textbf{Row 2:} $\rho_{\tilde{E}\tilde{T}}$ is fixed.
        \textbf{Row 3:} $\rho_{\tilde{E}\tilde{T}}$ and $\sigma_{\tilde{T}}$ are fixed.
        \textbf{Row 4:} $\rho_{\tilde{E}\tilde{T}}$, $\sigma_{\tilde{T}}$, and $\mu_{\tilde{T}}$ are fixed.
        Progressively constraining these parameters reduces degeneracies in the inferred intrinsic distributions and yields close agreement with the MCD measured shapes (Fig.~\ref{measured 3D shapes}).}
        \label{inferred 3D shapes}
    \end{figure*}

    \begin{table*}[]
        \centering
        \caption{Our inferred 3D shapes based on PySersic measurements}
        \label{inferred shape parameters}
        \resizebox{\textwidth}{!}{%
        \begin{tabular}{ccccccccccc}
        \hline\hline
        $z$ & $\log{M_\star/M_\odot}$ & $\mu_{\tilde{E}}$ & $\mu_{\tilde{T}}$ & $\mu_{\log{A}}$ & $\sigma_{\tilde{E}}$ & $\sigma_{\tilde{T}}$ & $\sigma_{\log{A}}$ & $\rho_{\tilde{E}\tilde{T}}$ & $\rho_{\tilde{E}L}$ & $\rho_{\tilde{T}L}$ \\
        \hline
        6.3-8.0 & 6.5-7.5 &  0.374$\pm$0.048 & 1.784$\pm$0.287 &
        -0.110$\pm$0.006 & 0.456$\pm$0.024 & 1.406$\pm$0.175   &
        0.213$\pm$0.004 & 0.601$\pm$0.186 &
        0.350$\pm$0.056 & 0.117$\pm$0.085 \\
        
        6.3-8.0 & 7.5-9.5 &  0.438$\pm$0.061 & 1.388$\pm$0.295 &
        -0.011$\pm$0.006 & 0.499$\pm$0.025 & 1.301$\pm$0.183   &
        0.213$\pm$0.005 & 0.486$\pm$0.208 &
        0.519$\pm$0.050 & 0.293$\pm$0.078 \\
        
        9.0-15.0 & 6.5-7.5 &  0.498$\pm$0.049 & 0.865$\pm$0.174 &
        -0.335$\pm$0.009 & 0.511$\pm$0.034 & 0.774$\pm$0.138   &
        0.230$\pm$0.007 & -0.115$\pm$0.216 &
        0.712$\pm$0.044 & -0.267$\pm$0.149 \\

        9.0-15.0 & 7.5-9.5 &  0.523$\pm$0.088 & 1.638$\pm$0.416 &
        -0.220$\pm$0.015 & 0.425$\pm$0.067 & 1.009$\pm$0.279   &
        0.200$\pm$0.014 & 0.022$\pm$0.293 &
        0.355$\pm$0.149 & -0.069$\pm$0.222 \\
        \hline
        \end{tabular}}
        \tablefoot{
          $L\equiv \log A$. Values are reported as posterior medians with $1\sigma$ uncertainties (16th--84th percentiles) from the HMC samples.
          }
    \end{table*}

    \subsection{Inferred 3D shapes}
    \label{section 3.3: Inferred 3D shapes}
    
    We now present intrinsic 3D shape distributions inferred from the projected measurements using the $\ltETa$ model introduced in Sect.~\ref{section 2.3: Inferring 3D Shapes}. 
    As discussed in Sect.~\ref{section 2.3.3: PyMC setup and tests} and demonstrated in Appendix~\ref{Appendix A}, parameters controlling the triaxiality dimension are typically weakly constrained for the high-redshift sample sizes available here ($N \lesssim 10^3$), in particular the mean of $\tilde{T}$ and its correlation with $\tilde{E}$.
    To illustrate the impact of these degeneracies, we perform the inference in a sequence of runs in which we progressively fix selected triaxiality-related parameters to the values measured directly from the star-particle distributions (i.e. the particle-measured constraints shown in Fig.~\ref{marginal fitted distribution}). 
    Operationally, this procedure corresponds to imposing increasingly informative priors guided by the simulation, and it isolates which components of the intrinsic covariance structure are most difficult to recover from projected data alone.  

    Figure~\ref{inferred 3D shapes} shows the resulting inferred intrinsic distributions in the $(C/A,\,B/A)$ plane. 
    From top to bottom, the rows correspond to: (i) a fully unconstrained run in which all nine parameters of the Gaussian in $(\tilde{E},\tilde{T},\log A)$ are free; (ii) a run with $\rho_{\tilde{E}\tilde{T}}$ fixed; (iii) a run with both $\rho_{\tilde{E}\tilde{T}}$ and $\sigma_{\tilde{T}}$ fixed; and (iv) a run with $\rho_{\tilde{E}\tilde{T}}$, $\sigma_{\tilde{T}}$, and $\mu_{\tilde{T}}$ fixed. 
    Across bins, the unconstrained model typically recovers the qualitative location of the intrinsic distribution, but artificial broadening can appear when the triaxiality direction remains poorly constrained. 
    Fixing $\rho_{\tilde{E}\tilde{T}}$ already stabilizes the solution in the most degenerate cases, and additionally fixing $\sigma_{\tilde{T}}$ and $\mu_{\tilde{T}}$ brings the inferred distributions into substantially closer agreement with the particle-measured shapes.
    While Fig. \ref{inferred 3D shapes} presents the marginalized constraints on our shape parameters, a complete visualization of the multidimensional posterior distributions, confirming the stability and unimodality of our inference, is provided in Appendix \ref{Appendix B}.
    
    These results underscore that, for $N\lesssim10^3$, projected $(b/a,\log a)$ data alone often provide limited leverage on the triaxiality dimension (see Appendix~\ref{Appendix A}). 
    At the same time, they demonstrate that the $\ltETa$ model is sufficiently flexible to reproduce the asymmetric and skewed intrinsic-shape distributions of high-redshift dwarf galaxies (see Appendix \ref{Appendix C} for quantified comparison), when additional information is available, either through larger samples or through informative priors. 
    A direct comparison to the standard $\ETa$ model is provided in Appendix~\ref{Appendix D}.

    \subsection{Sample size and HMC model choice}
    We refine the interpretation of our results as follows. 
    On the \FIRE dwarf galaxy population, forward-modeling Bayesian inference is able to reproduce the intrinsic 3D shape distribution measured directly from star particles, demonstrating that the overall methodology is robust when its intrinsic-population model is well matched to the data and when the relevant parameters are identifiable at the available sample size. 
    In particular, once the intrinsic distribution is parameterized in a form that respects the bounded nature and skewness of the shape parameters (for our sample, the \textit{ltETlogA} model, i.e., a Gaussian in $(\tilde{E},\tilde{T},\log A)$, equivalent to JohnsonSB in $(E,T)$), the inferred $(C/A,\,B/A)$ distributions closely track the particle-measured “ground truth”.
    
    A central limitation is not the forward-modeling machinery itself, but the difficulty in inferring triaxiality-related parameters for comparably modest (here $N\lesssim 10^3$) sample sizes. 
    This manifests most directly in the correlations involving triaxiality (e.g. $\rho_{\tilde{E}\tilde{T}}$) and, in some bins, in the mean of $\tilde{T}$. 
    We therefore use the simulation to quantify the sample size required to fit the fully general covariance structure (i.e. allowing all correlations), and we show that substantially larger samples ($N\sim10000$, see Appendix \ref{Appendix A}) are needed before all triaxiality-related parameters can be recovered without informative priors. 
    Conversely, when those parameters are fixed or strongly regularized, as a stand-in for external information from larger surveys or simulation-based priors, the inferred intrinsic distributions converge to the measured ones.
    
    Finally, our results emphasize that the appropriate intrinsic-distribution family is population-dependent. 
    The standard Gaussian assumption in $(E,T,\log A)$ may provide a reasonable approximation for many populations, but it is not universally valid. 
    For the high-$z$ dwarf population studied here, the intrinsic distributions of $(E,T)$ are better described by JohnsonSB marginals, motivating the \textit{ltETlogA} parameterization. 
    In this sense, our work does not argue for a single universal model, but rather provides (i) a validated alternative distributional choice, and (ii) an explicit assessment of the sample-size requirements for adopting a fully general model with all correlations.

%%%%%%%%%%%%%%%%%%%%%%%%%%%%%%%%%%%%%%%%%%%%%%%%%%%%%%%%%%%%%%
\section{Discussion}
\label{section 4: Discussion}
    \subsection{Implications from \FIRE Simulation}
    In our analysis of the \FIRE simulation, we find that the intrinsic 3D shapes of low-mass galaxies ($\Mstar=6.5$ -- $9.5$) at $z > 6.3$ are predominantly prolate (triaxial) and spheroidal. Conversely, strongly oblate galaxies are rarely inferred in this regime. These findings are consistent with the broader consensus of other cosmological and zoom-in simulation studies. For instance, analyses of the VELA \citep{Ceverino2015, Zhang2019}, and TNG50 \citep{Pillepich2019} similarly report that high-redshift, low-mass galaxies tend to be strongly prolate. 
    The convergence of these independent simulations suggests that prolate geometries are a fundamental physical characteristic of low-mass galaxies during the Epoch of Reionization under current galaxy formation paradigms, rather than a statistical artifact of any single model.
    
    Recent morphological analyses of early JWST CEERS data \citep{Pandya2024} find that the galaxies observed are either disks or prolate.
    While their inference of highly elongated structures aligns broadly with our findings, their conclusion that these systems possess a significant oblate component stands in tension with our simulated spheroidal population. 
    It is important to emphasize, however, that the redshift distribution of the CEERS observational sample differs from our simulated high-redshift ($z > 6.3$) sample, which may partially account for this morphological discrepancy.
    
    Alternatively, intrinsic shapes can be constrained dynamically via stellar kinematics, which avoids the projection degeneracies inherent to photometric studies. 
    Recent work by \citet{Wang2026} uses stellar kinematics to discriminate between edge-on oblate discs and side-on prolate spindles. 
    Their results support the prolate spindle scenario, which is in good agreement with both our simulation predictions and current photometric observations, although the physical nature of the remaining secondary population remains unclear. 
    We also note that their kinematic sample only extends up to $z \approx 4$. Consequently, bridging this gap with deeper, higher-resolution spectroscopic and imaging surveys at $z > 6$ will be required to definitively resolve these morphological tensions and fully map the structural evolution of early galaxies.

    \subsection{Implications for the Inference Methodology}
    From a methodological perspective, forward-modeling inference approaches often rely on several simplifying assumptions: (1) a single morphological population is sufficient to describe the entire dataset, (2) the intrinsic shape and size parameters, usually $(E, T, \log A)$, follow a multivariate normal distribution, and (3) only the correlation between intrinsic ellipticity and physical size needs to be explicitly modeled. Our direct 3D measurements from the \FIRE simulation allow us to evaluate the validity of these assumptions for early, low-mass galaxies.
    
    First, our analysis demonstrates that assuming a single underlying morphological population is statistically robust when modeling low-mass galaxies at $z > 6.3$. Because these early systems are overwhelmingly triaxial and prolate, introducing multiple populations is unnecessary. 
    However, this assumption is highly sensitive to the stellar mass and cosmic epoch being studied. 
    For instance, \citet{Zhang2022} applied a similar forward-modeling framework to local galaxies ($z < 0.9$) in the KiDS+VIKING surveys across a higher mass range of $\log(M_*/M_\odot) = 9.0$ -- $11.5$. 
    They found that using a double multivariate normal distribution significantly improved the fits to the observed projected axis ratio and size distributions. 
    While their work focused on quiescent galaxies, it highlights a broader necessity: as the universe evolves and galaxies diversify into co-existing populations of disks and spheroids, multi-modal inference models may become essential.

    Secondly, we find that the data are better described by a multivariate normal distribution in logit-space ($\logit{}E, \logit{}T, \log{}A)$ rather than in $(E, T, \log A)$ space. 
    The logit-based distribution naturally accommodates the strongly triaxial and spheroidal geometries characteristic of these early dwarfs and limits the possible ranges of $E, T$ to fall between 0 and 1. 
    Given that other cosmological simulations, such as VELA\citep{Zhang2019} and TNG50\citep{Pillepich2019}, predict morphological distributions similar to \FIRE, we anticipate that this transformed parameterization will be broadly applicable across theoretical models. 

    \subsection{Caveats}
    While our forward-modeling framework provides robust constraints on the intrinsic shapes of high-redshift dwarf galaxies, there are several caveats and simplifying assumptions in our methodology that warrant discussion.
    
    First, our intrinsic 3D shape measurements using the Minimum Covariance Determinant (MCD) estimator are fundamentally based on the spatial distribution of stellar mass (i.e., star particles). 
    However, observational shape measurements such as projected Sérsic profile fitting, are inherently sensitive to surface brightness rather than mass. 
    To make the most direct comparison with observations, future studies should consider employing a luminosity-weighted 3D shape estimator \citep{klein2026}. 
    A luminosity-weighted approach would naturally account for the outsized influence of young, bright star-forming regions on the observed geometry, which could differ slightly from the underlying stellar mass distribution.
    
    Second, when generating the mock JWST observations, our radiative transfer modeling with SKIRT has generic uncertainties arising from, for example, dust size distribution, IMF, and dust-to-metal ratio.
    We note, however, that low-mass dwarf galaxies at high redshift are generally expected to be dust poor.
    Hence, we do not expect these uncertainties to have a significant impact on our findings.
    We also note that the random noise added to our mock JWST observations is suppressed and various systematics from realistic images are not accounted for.
    
    Third, we must acknowledge the observational limits imposed by the Point Spread Function (PSF). 
    Many of the simulated dwarf galaxies in our sample possess intrinsic physical sizes that fall below the full width at half maximum (FWHM) of the JWST NIRCam PSF. 
    In our analysis, this does not bias our inferred shape distributions because we use the mock images generated using the exact, known simulated PSF and later de-convolved to estimate the apparent shape parameters, which is used in the Bayesian inference approach. 
    However, when applying this framework to real observational data, where the PSF is empirically reconstructed and subject to spatial and temporal variations.
    Reliably extracting the axis ratios of such ultra-compact, sub-PSF systems will introduce an additional layer of systematic uncertainty.
    
    Finally, there are important statistical limitations tied to our sample sizes. As argued previously, fully capturing the complex geometry of these systems requires modeling the full correlation matrix between the shape and size parameters. 
    However, as demonstrated in Appendix\ref{Appendix A}, fitting a full covariance matrix in this high-dimensional parameter space is highly susceptible to model degeneracies unless it is tightly constrained by a large sample of galaxies. This limitation severely impacts our highest redshift bin ($z = 9-15$); the paucity of simulated galaxies at these extreme redshifts within the \FIRE volume prevents us from making statistically reliable inferences for this specific epoch (see Fig. \ref{2d shape properties}). 
    Future work leveraging larger-volume cosmological simulations will be essential to provide the sample sizes necessary to break these degeneracies and securely map the morphological evolution of galaxies into the cosmic dawn.

%%%%%%%%%%%%%%%%%%%%%%%%%%%%%%%%%%%%%%%%%%%%%%%%%%%%%%%%%%%%%%
\section{Conclusions}
\label{section 5: Conclusion}
    In this paper we investigate the projected sizes and morphologies of galaxies within the first billion years of cosmic time using the \FIRE simulation suite \citep{Feldmann2025}. 
    We generate JWST-like mock observations for halos spanning $z=6.3$ -- $15$ and $\Mstar=6.5$ -- $9.5$, adopting negligible background noise to isolate methodological systematics. 
    From these mocks we select 2366 high-quality sources and divide them into four mass--redshift bins defined by $\Mstar\in[6.5,7.5],[7.5,9.5]$ and $z\in[6.3,8],[9,15]$. 
    For each galaxy we measure projected half-light radii and axis ratios by fitting single-component S\'ersic models to the mock images using \texttt{PySersic} \citep{Pasha2023}.
    
    To assess the robustness of forward-modeling methods used for observational samples, we inferred the intrinsic 3D shape distribution from the observed joint distribution of projected axis ratios and sizes, $(q,\log a)$, using a Bayesian framework implemented in \texttt{PyMC} \citep{salvatier2015}. 
    We compare the inferred intrinsic distributions to ``ground-truth'' intrinsic shapes measured directly from star-particle data using an outlier-resistant Minimum Covariance Determinant (MCD) estimator \citep{Rousseeuw1984}, approximating each galaxy as a triaxial ellipsoid with principal axes $(A,B,C)$. 
    This comparison allowed us to test the key assumptions entering recent JWST-based 3D-shape inference analyses.
    
    Our main results are:
    
    \begin{enumerate}
    \item
    Across all four mass--redshift bins, the projected axis-ratio distribution peaks at $q\simeq 0.4$--$0.6$. 
    Comparing to the Dawn JWST Archive catalog \citep{Genin2025}, we find good agreement in the $q$ distributions (Fig.~\ref{2d shape properties}). 
    The size distributions show a similar shape as the observed ones, but the simulated galaxies are systematically larger by $\simeq 0.3$\,dex (a factor of $\sim 2$) in projected size. 
    
    \item
    The MCD-based measurements show that the \FIRE dwarfs are predominantly elongated (prolate/triaxial) with a non-negligible spheroidal component, and only a small fraction of clearly oblate systems (Fig.~\ref{measured 3D shapes}). 
    This picture is consistent with VELA, TNG-50 \citep{Zhang2019,Pillepich2019} simulations.
    The intrinsic marginal distributions of ellipticity and triaxiality are well described by JohnsonSB distributions, implying that the transformed variables $(\logit{E},\logit{T})$ are Gaussian (Fig.~\ref{marginal fitted distribution}). 
    This motivates modeling the intrinsic population as a multivariate normal distribution in $(\logit{E},\logit{T},\log A)$ rather than in $(E,T,\log A)$, and we find that allowing correlations involving triaxiality is necessary for a flexible description of the intrinsic covariance structure.
    
    \item
    Using the simulation as a controlled testbed, we find that forward-modeling inference can reproduce the particle-measured intrinsic distributions for this population when the intrinsic-population model is appropriately specified and when the relevant parameters are identifiable. 
    In particular, the standard Gaussian model in $(E,T,\log A)$ ($\ETa$) can yield biased intrinsic distributions for the \FIRE dwarfs, whereas our logit-transformed model ($\ltETa$) successfully recovers the MCD-measured shapes (Fig.~\ref{inferred 3D shapes}; Appendix \ref{Appendix D}). 
    A dominant limitation is the weak constraining power on triaxiality-related parameters from projected data alone with sample sizes $N\lesssim 10^3$: correlations involving $\logit{T}$ (or $T$) and, in some bins, the mean/scatter of $\logit{T}$ (or $T$) remain poorly constrained. 
    We show that adopting informative priors (or equivalently fixing the least constrained triaxiality parameters to externally motivated values) stabilizes the inference and yields close agreement with the particle-measured distributions.
    \end{enumerate}
    
    Overall, our results support forward-modeling Bayesian inference as a robust tool for constraining the galaxy 3D structure of low-mass galaxies at high redshift, provided that (i) the intrinsic-population model is appropriate for the galaxy population under study and (ii) sample sizes are sufficiently large or a sufficiently informative prior on the triaxiality parameters is available. 

%%%%%%%%%%%%%%%%%%%%%%%%%%%%%%%%%%%%%%%%%%%%%%%%%%%%%%%%%%%%%%

%%%%%%%%%%%%%%%%%%%%%%%%%%%%%%%%%%%%%%%%%%%%%%%%%%%%%%%%%%%%%%
\begin{acknowledgements}
      We thank Haowen Zhang and Oliver Hahn for useful discussions, and Viraj Pandya for providing their code used in \citet{Pandya2024}.
      This research was supported in part by grant NSF PHY-2309135 to the Kavli Institute for Theoretical Physics (KITP).
      This work was performed in part at the Aspen Center for Physics, which is supported by National Science Foundation grant PHY-2210452.
      This work was supported in part by an allocation from the Swiss National Supercomputing Centre (CSCS) under project IDs s697 and s698. We acknowledge access to Piz Daint and Alps/Eiger at the Swiss National Supercomputing Centre, Switzerland, under the University of Zurich’s share with the project ID uzh18. This work made use of infrastructure services provided by S3IT (www.s3it.uzh.ch), the Service and Support for Science IT team at the University of Zurich.
      JSB acknowledges support from NSF grant AST-2408246.
      XS acknowledges the support from NASA theory grant JWST-AR-04814.
      This work made use of the Python packages \texttt{PySersic} \citep{Pasha2023}, \texttt{PyMC} \citep{salvatier2015}, \texttt{numpy} \citep{Numpy2020}, \texttt{scipy} \citep{Scipy2020}, \texttt{matplotlib} \citep{Hunter2007}, \texttt{corner} \citep{corner2016}, \texttt{Linmix} \citep{Kelly2007}.
\end{acknowledgements}

\section*{Data availability}
This paper is based on the galaxy catalog from the DAWN JWST Archive (DJA) \citep{Genin2025}, which is publicly available at \url{https://dawn-cph.github.io/dja/}. The data from the \FIRE simulation, as well as the plotting scripts and analysis codes used in this work, are available from the corresponding author upon reasonable request.

% Please note that we have included the references below in
% order to compile the document, but we ask you to:
\bibliographystyle{bibtex/aa}
\bibliography{reference}

%%%%%%%%%%%%%%%%%%%%%%%%%%%%%%%%%%%%%%%%%%%%%%%%%%%%%%%%%%%%%%%
% Appendices must be placed after   \end{thebibliography}
% They will be placed automatically on a new page.
%%%%%%%%%%%%%%%%%%%%%%%%%%%%%%%%%%%%%%%%%%%%%%%%%%%%%%%%%%%%%%%
\begin{appendix}

\twocolumn
\section{HMC inference test}
\label{Appendix A}

To validate the $\ltETa$ inference framework and to quantify sample-size requirements, we performed parameter-recovery tests following Appendix~A of \citet{Pandya2024}. 
We adopt a set of nine ``true'' population parameters $(\vec{\mu},\Sigma)$ measured from the MCD-based intrinsic shapes of the Fire1-dominated population (Fig.~\ref{marginal fitted distribution}). 
Using these parameters, we generate the model prediction for the projected $(q,\log a)$ distribution, and then draw Poisson realizations with total sample sizes $N\in\{300,1000,3000,10^4,3\times10^4\}$ to mimic mock observational samples.

\begin{figure*}
    \centering
    \includegraphics[width=\textwidth]{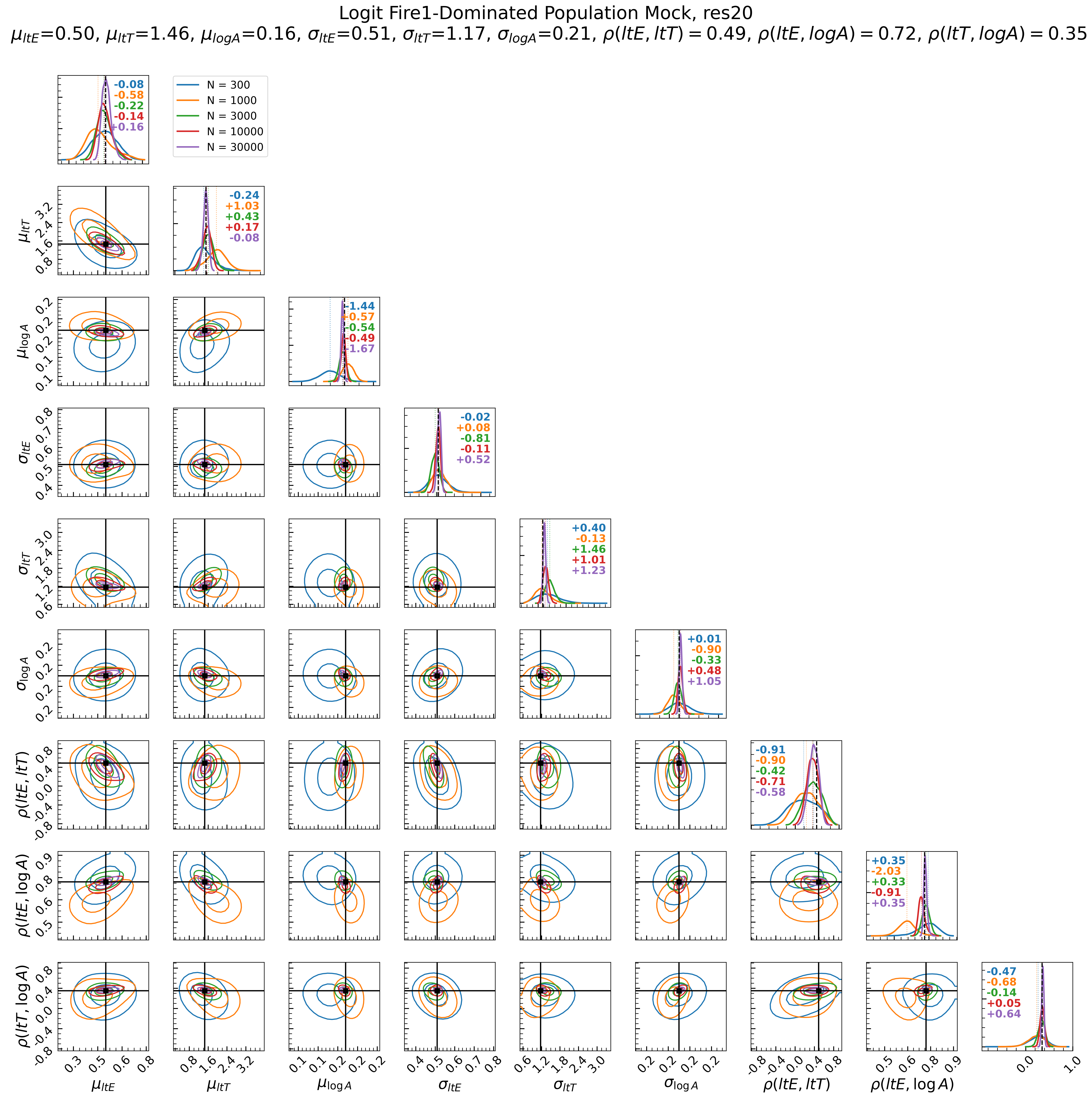}
    \caption{
    Parameter-recovery test for the \textit{ltETlogA} model using mock samples drawn from a known ``true'' population (Fire1-dominated; values listed above the plot). 
    Curves/contours show posterior constraints for sample sizes $N=300,1000,3000,10^4,$ and $3\times10^4$ at a $(q,\log a)$ histogram resolution of 20 bins per axis. 
    Black lines mark the true input values. 
    Off-diagonal contours enclose the 39\% and 86\% highest-posterior-density regions (``$1\sigma$'' and ``$2\sigma$'' for a bivariate Gaussian). 
    Coloured annotations in the diagonal panels report the normalized bias $\Delta$ (difference between posterior median and truth in units of the posterior standard deviation). 
    Most parameters tighten rapidly with increasing $N$, whereas $\rho_{\tilde{E}\tilde{T}}$ remains comparatively weakly constrained even at large $N$, reflecting a degeneracy in the projected-data summary.
    }
    \label{Fig A: HMC test fire1 ltETa0.14 res20}
\end{figure*}

Figure~\ref{Fig A: HMC test fire1 ltETa0.14 res20} shows an example recovery experiment using a $(q,\log a)$ histogram resolution of 20 bins per axis. 
The diagonal panels show the one-dimensional marginal posteriors for each parameter. 
The off-diagonal panels show posterior credible regions in two dimensions; the contours labeled ``$1\sigma$'' and ``$2\sigma$'' correspond to the 39\% and 86\% enclosed-probability regions for a bivariate Gaussian, respectively. 
The black reference lines indicate the true input values. 
In the diagonal panels, the colored annotations report the normalized bias,
\begin{equation}
\delta \equiv \frac{\mathrm{median}(\mathrm{posterior}) - \mathrm{truth}}{\sigma_{\mathrm{posterior}}},
\end{equation}
for each sample size.

The recovery tests show that most location and scale parameters in $(\tilde{E},\tilde{T},\log A)$ become relatively well constrained once $N\gtrsim 10^3$, while at $N\sim 300$ the posteriors remain substantially broader but are typically consistent with the truth within uncertainties. 
In contrast, the correlation coefficient $\rho_{\tilde{E}\tilde{T}}$ is only weakly constrained even for very large $N$, and its posterior remains comparatively broad at $N=3\times10^4$. 
This behavior indicates an intrinsic identifiability limitation: for this projected-data summary (binned $(q,\log a)$ counts), variations in $\rho_{\tilde{E}\tilde{T}}$ can be partially compensated by changes in other parameters, yielding similar projected distributions.

We also tested the interplay between statistical and discretization errors. 
The histogram resolution and the discreteness of the precomputed ellipsoid library set a systematic floor, whereas increasing $N$ decreases statistical uncertainties. 
If the library/grid is too coarse, the posterior can become artificially tight around a biased value once $N$ is sufficiently large. 
For the libraries used in this work (grid step $\Delta=0.28$, $2.5\times10^5$ ellipsoids; and $\Delta=0.14$, $10^6$ ellipsoids), we find no apparent differences in the recovered parameters when varying the histogram resolution from 20 to 40 bins over the range of $N$ explored here.

These controlled experiments support the main conclusion of the paper: forward-modeling inference can recover the intrinsic population distribution for \FIRE-like dwarfs when the intrinsic model is appropriate and the sample size is sufficient, but triaxiality-related correlations (especially $\rho_{\tilde{E}\tilde{T}}$) can remain weakly constrained without informative priors or external constraints.

%%%%%%%%%%%%%%%%%%%%%%%%%%%%%%%%%%%%%%%%%%%%%%%%%%%%%%%%%
\section{\texorpdfstring{$\ltETa$}{ltETlogA} posterior results}
\label{Appendix B}

\begin{figure*}
    \centering
    \includegraphics[width=\textwidth]{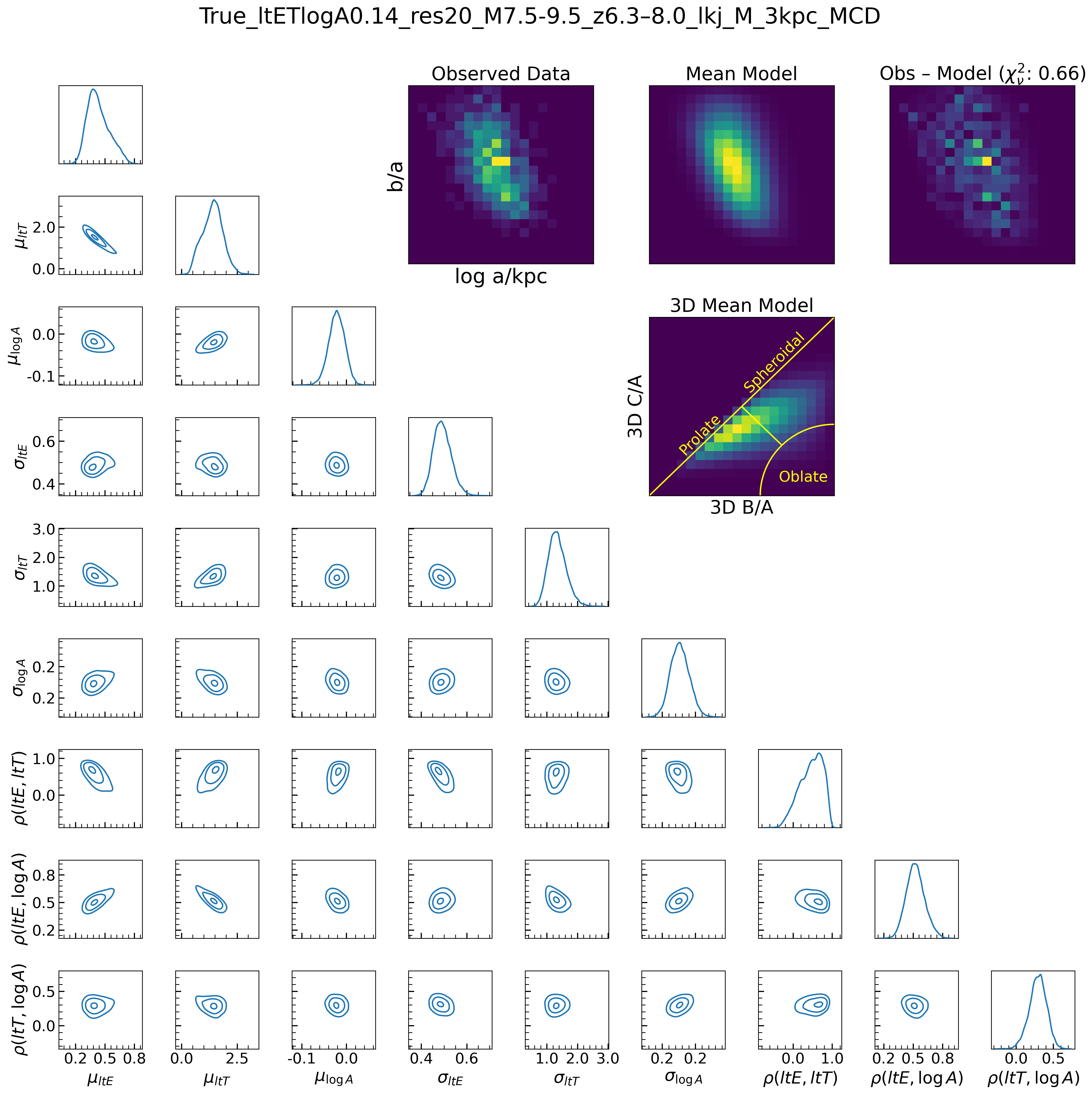}
    \caption{
    Full posterior corner plot for the $\ltETa$ inference model applied to the z = 6.3--8.0, $\log(M_\star/M_\odot)$ = 7.5--9.5 bin. 
    Diagonal panels show 1D marginal distributions for the nine model parameters (means, standard deviations, and correlation coefficients in the transformed space), while off-diagonal panels display 2D credible regions. 
    The inset panels in the upper right present a posterior predictive check: "Observed Data" shows the input mock projected measurements in the (b/a, $\log a$) plane, "Mean Model" shows the projection generated from the posterior mean parameters, and "Obs - Model" displays the residual with a reduced chi-square ($\chi^2_\nu$) of 0.66. 
    The "3D Mean Model" panel illustrates the corresponding reconstructed intrinsic shape distribution in the physical (C/A, B/A) plane.
    }
    \label{Fig B: HMC inference posterior}
\end{figure*}

To fully characterize the parameter space and evaluate the stability of our Bayesian framework, we examine the full posterior distributions of the $\ltETa$ model. 
The corner plot in Fig. \ref{Fig B: HMC inference posterior} presents the results for the fully unconstrained run applied to the z = 6.3–8.0, $\log(M_\star/M_\odot)$ = 7.5–9.5 mass-redshift bin. 
The diagonal panels confirm that the location and scale parameters in the logit-transformed space, specifically the means ($\mu$) and standard deviations ($\sigma$) for ellipticity, triaxiality, and size, are well constrained by the mock observational data, showing well-behaved, unimodal marginal distributions.
As discussed in Section 3.4, the correlation coefficients, particularly between ellipticity and triaxiality, exhibit broader posteriors, reflecting the inherent degeneracies when recovering 3D triaxiality from 2D projected data alone.

The panels in the upper right quadrant of Fig. \ref{Fig B: HMC inference posterior} serve as a visual posterior predictive check in both the projected observation space and the intrinsic physical space. 
The "Observed Data" panel displays the 2D histogram of the mock projected measurements for this specific bin in the (b/a, $\log a$) plane. 
Adjacent to it, the "Mean Model" panel shows the theoretical projected distribution reconstructed using the mean values of the inferred posterior parameters. 
The "Obs - Model" panel maps the residual difference between the two, reporting a reduced chi-square ($\chi^2_\nu$) of 0.66, which indicates a robust fit that captures the data well without significant overfitting. 
Finally, the "3D Mean Model" panel visualizes the corresponding intrinsic shape distribution mapped back into the physical (C/A, B/A) plane, highlighting the dominant prolate and spheroidal geometries inferred by the model.

%%%%%%%%%%%%%%%%%%%%%%%%%%%%%%%%%%%%%%%%%%%%%%%%%%%%%%%%%
\section{Statistical analysis of \texorpdfstring{$\ltETa$}{ltETlogA} and MCD results}
\label{Appendix C}

\begin{figure*}
    \centering
    \includegraphics[width=\textwidth]{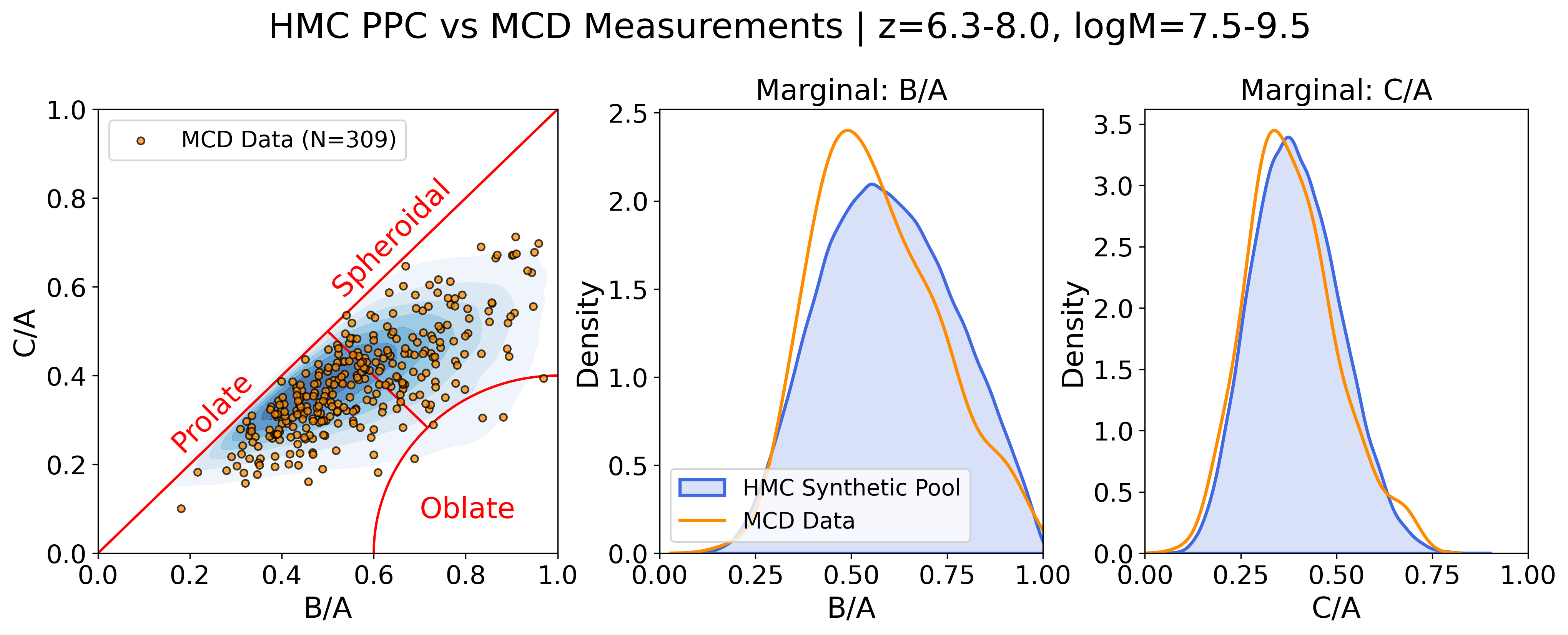}
    \caption{
    Comparison between the intrinsic 3D shapes recovered by the $\ltETa$ HMC inference (Synthetic Pool) and the direct robust measurements from star particles (MCD Data) for the z = 6.3--8.0, $\log(M_\star/M_\odot)$ = 7.5--9.5 bin. 
    The left panel shows the 2D distribution overlaid in the (C/A, B/A) shape plane, with classification boundaries marked in red. 
    The middle and right panels display the 1D marginal density distributions for the intermediate-to-major (B/A) and minor-to-major (C/A) axis ratios, respectively, demonstrating reasonable morphological agreement between the inferred model and the ground-truth stellar distributions.
    }
    \label{Fig C: HMC inference Comparison}
\end{figure*}

To validate our forward-modeling framework, we evaluate whether the inferred 3D intrinsic shape distributions are structurally consistent with the true underlying simulated populations. 
We are validating an inverse problem that relies on lossy and approximate forward modeling.
Therefore, the core question we aim to address is: Is the measured intrinsic data a typical realization compatible with the observed projected data under our assumed model?

\subsection{Formalism and Posterior Predictive Check}
Our validation procedure can be formalized as follows. 
Let our complex, true 3D spatial data (the star particles) be $D = \{x_n\}_{n=1}^N$. We compress this data into intrinsic shape summaries via a shape estimator $S$, such that:
\begin{equation}
    x'_n = S(x_n), \quad D' = \{x'_n\}, \quad x'_n \sim p(x' \mid \theta),
\end{equation}
where $\theta$ represents the population-level intrinsic shape parameters. The observed data are 2D projections of the complex data, defined as:
\begin{equation}
    x''_n = F(x_n), \quad D'' = \{x''_n\}.
\end{equation}
In our Bayesian inference framework, we rely on an analytical approximation for the projection, assuming $F(x) \approx f(S(x)) = f(x')$. Using this forward model, we infer the posterior distribution of the population parameters:
\begin{equation}
    \theta^{(k)} \sim p(\theta \mid D'').
\end{equation}

To test whether the inferred posterior is consistent with the true intrinsic data $D'$, we perform a posterior predictive check (PPC) directly in the intrinsic space. The predictive distribution is approximated by marginalizing over the posterior:
\begin{equation}
    p(D' \mid D'') = \int p(D' \mid \theta) p(\theta \mid D'') \, d\theta.
\end{equation}
Operationally, we draw parameters from the posterior $\theta^{(k)}$ and simulate replicate intrinsic datasets $D_{\mathrm{rep}}^{\prime (k)} \sim p(D' \mid \theta^{(k)})$. We then compare the true data to the replicates using discrepancy measures $d(\cdot,\cdot)$, specifically the Maximum Mean Discrepancy (MMD, \citet{gretton2012}) for multivariate comparisons and the 2D Wasserstein distance (W2D, \citet{villani2009}, \citet{peyre2019}) on derived quantities. 

We compute the distance between the observed intrinsic data and the replicates:
\begin{equation}
    \Delta_{\mathrm{obs}}^{(k)} = d\big(D', D_{\mathrm{rep}}^{\prime (k)}\big)
\end{equation}
and calibrate this against replicated--replicated distances to establish a baseline for expected statistical fluctuations:
\begin{equation}
    \Delta_{\mathrm{rep}}^{(k)} = d\big(D_{\mathrm{rep}}^{\prime (k)}, \tilde{D}_{\mathrm{rep}}^{\prime (k)}\big).
\end{equation}
The consistency is then evaluated via the PPC $p$-value:
\begin{equation}
    p_{\mathrm{PPC}} = \Pr\big(\Delta_{\mathrm{rep}} \ge \Delta_{\mathrm{obs}} \mid D''\big).
    \label{Eq C.7}
\end{equation}

\subsection{Interpretation of the Statistical Results}
To provide a visual context for the posterior predictive checks, Fig \ref{Fig C: HMC inference Comparison} compares the intrinsic 3D shapes recovered by the $\ltETa$ HMC inference (Synthetic Pool) to the direct robust measurements from star particles (MCD Data) for the $z=6.3-8.0$, $\log(M_{*}/M_{\odot})=7.5-9.5$ bin.
The statistical results of this comparison for our fiducial model reveal an important nuance in evaluating forward-modeled shapes. 
For this sample bin, the Bayesian Posterior Predictive p-values (Equation \ref{Eq C.7}) are 0.014 for MMD and 0.024 for W2D, representing a single global probability across the posterior ensemble.
In contrast, evaluating the 1D marginals yields a distribution of p-values, as we compute individual Kolmogorov-Smirnov (KS) tests per posterior draw, allowing us to report median p-values with their $1\sigma$ (16th–84th percentile) uncertainties.
Given that we are ultimately comparing two fundamentally different quantities, namely measurements derived from projected 2D light distributions versus those derived directly from 3D stellar mass distributions, it is plausible that the morphological distributions look remarkably similar by eye, although they exhibit formally significant differences when subjected to rigorous statistical analysis (0.01 < p < 0.05). 

In a strict statistical sense, a median p-value < 0.05 indicates that the inferred distribution is formally distinguishable from the exact star-particle distribution.
This implies that the forward-modeling procedure does not perfectly capture every statistical subtlety of the true intrinsic shapes. 
This is likely due to the slight breakdown of the approximation $f(S(x))\approx F(x)$ introduced by the differing definitions of half-mass versus half-light radii.  

However, while the model is not a perfect statistical replica, it remains highly effective as an astrophysical approximation, and the 1D marginals actually look good enough for recovering the core physical properties. 
This is evidenced by the 1D Kolmogorov-Smirnov (KS) p-values for the individual marginal axis ratios B/A and C/A. 
Although the median KS p-values are small, their $1\sigma$ upper bounds extend significantly above the 0.05 threshold (e.g., $0.024_{-0.024}^{+0.356}$ for B/A and $0.047_{-0.047}^{+0.427}$ for C/A). 
This wide positive skew demonstrates that a substantial fraction of the posterior parameter draws generate intrinsic populations whose marginal distributions are statistically indistinguishable from the simulated data.  
Therefore, while minor systematic discrepancies exist, the $\ltETa$ inference successfully and robustly recovers the macroscopic morphological trends (i.e., the prolate-dominated geometry) of the target population.

% \begin{table*}[!h]
%     \centering
%     \caption{Statistical comparison between the 3D intrinsic shapes inferred via the $\ltETa$ model and direct MCD measurements from simulated star particles.}
%     \label{tab C: stats between the inference and measurement}
%     \resizebox{\textwidth}{!}{%
%     \begin{tabular}{ccccccccccc}
%     \hline\hlineC
%     $z$ & $\log{M_\star/M_\odot}$ & $MMD_{Model}$ & $MMD_{Base}$ & $p$-value MMD & $W2D_{Model}$ & $W2D_{Base}$ & $p$-value W2D & KS $p$-value $B/A$ & KS $p$-value $C/A$ \\
%     \hline
    
%     6.3-8.0 & 7.5-9.5 & $0.136_{-0.041}^{+0.078}$ & $0.045^{+0.022}_{-0.016}$ &
%     0.014 & $0.052^{+0.024}_{-0.013}$ & $0.027^{+0.006}_{-0.004}$ &
%     0.024 & $0.024^{+0.336}_{-0.024}$ & $0.047^{+0.427}_{-0.047}$ \\
    
%     \hline
%     \end{tabular}}
%     \tablefoot{
%       Values are reported as posterior medians with $1\sigma$ uncertainties (16th--84th percentiles) derived from the HMC samples. Model metrics refer to distances between the observed intrinsic data and posterior replicates, while Base metrics represent the calibrated statistical baseline from replicated-replicated distances.
%       }
% \end{table*}

%%%%%%%%%%%%%%%%%%%%%%%%%%%%%%%%%%%%%%%%%%%%%%%%%%%%%%%%%
\section{\texorpdfstring{$\ETa$}{ETlogA} model result}
\label{Appendix D}
\begin{figure*}
    \centering
    \includegraphics[width=\textwidth]{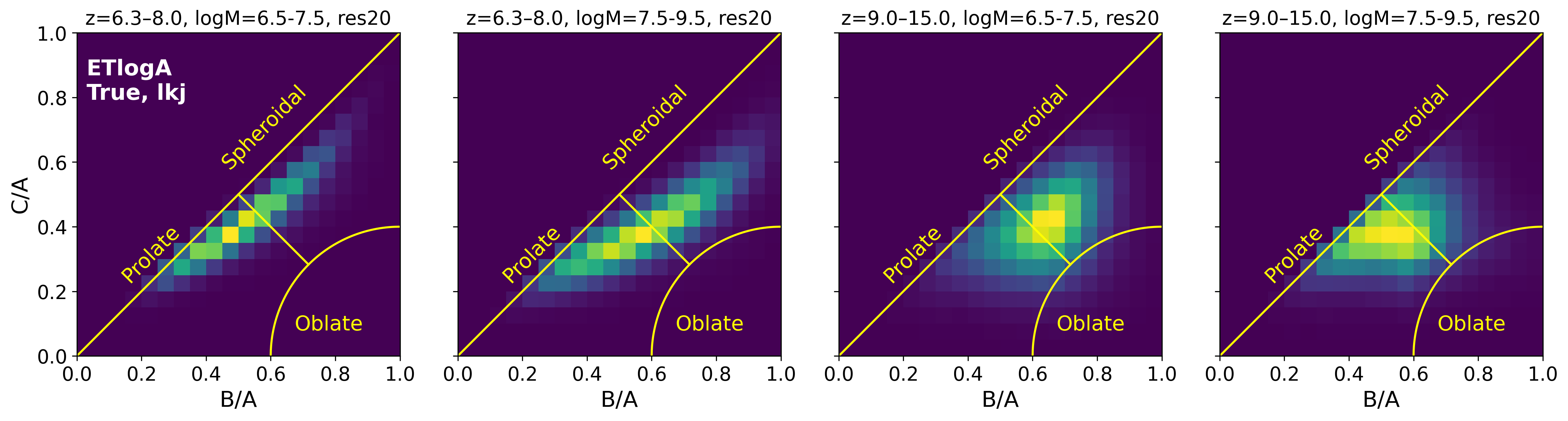}
    \includegraphics[width=\textwidth]{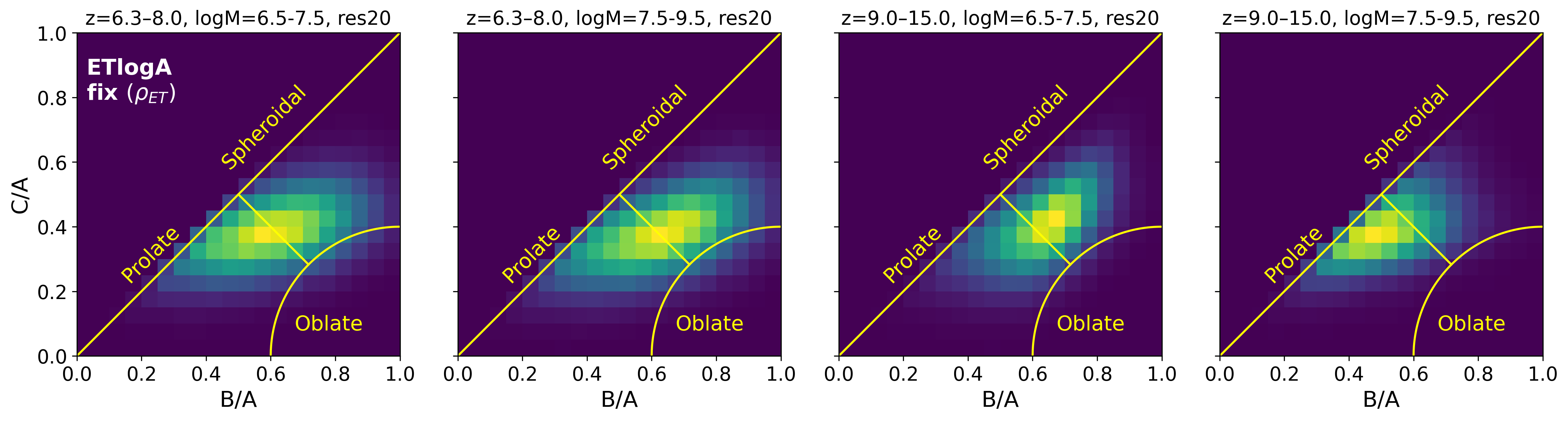}
    \includegraphics[width=\textwidth]{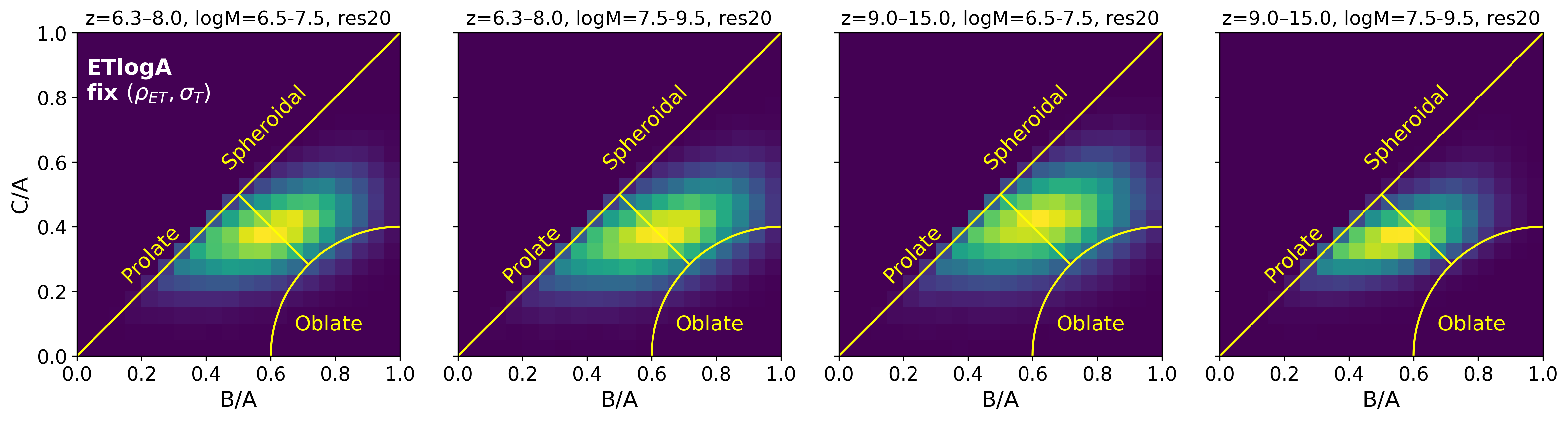}
    \includegraphics[width=\textwidth]{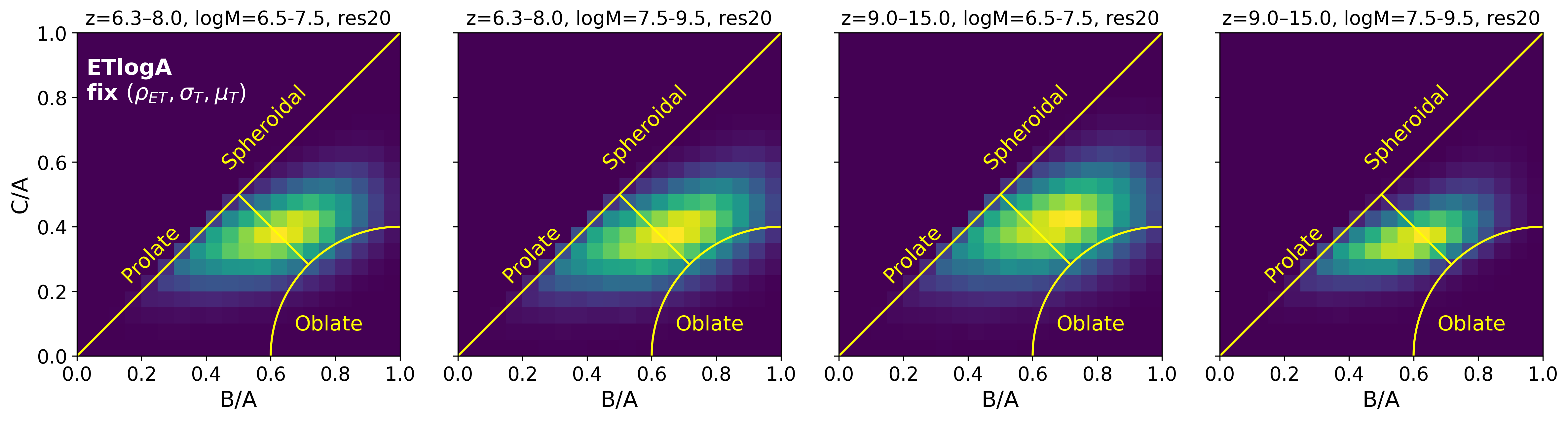}
    \caption{
    Inferred intrinsic 3D shape distributions in the $(C/A,\,B/A)$ plane using the standard \textit{ETlogA} model, shown in the same format as Fig.~\ref{inferred 3D shapes}. 
    Rows correspond to fixing triaxiality-related parameters to the particle-measured values: 
    (1) all parameters free; 
    (2) $\rho_{ET}$ fixed; 
    (3) $\rho_{ET}$ and $\sigma_T$ fixed; 
    (4) $\rho_{ET}$, $\sigma_T$, and $\mu_T$ fixed. 
    Compared to \textit{ltETlogA}, the \textit{ETlogA} model more readily produces extreme $\rho_{ET}$ values in unconstrained runs for the $z=6.3$--8 bins, whereas imposing the measured triaxiality constraints stabilizes the recovered intrinsic distributions.
    }
    \label{Fig D: ETa inferred 3D shapes}
\end{figure*}

Figure~\ref{Fig D: ETa inferred 3D shapes} shows the inferred intrinsic shape distributions obtained with the standard $ETa$ model, using the same sequence of ``fixing'' experiments as in Fig.~\ref{inferred 3D shapes}. 
The $\ETa$ model assumes a multivariate normal distribution directly in $(E,T,\log A)$, which implies an approximately symmetric behavior in the bounded variables $(E,T)$ once truncation is applied at the physical limits.

In the fully unconstrained runs (top row), the $\ETa$ model can favor very large values of $\rho_{ET}$ in the $z=6.3$--8 bins, effectively using an extreme ellipticity--triaxiality correlation to reproduce skewed intrinsic distributions that are better captured by the logit-Gaussian ($\ltETa$) parameterization. 
Fixing $\rho_{ET}$ to the particle-measured value largely removes this pathology and stabilises the inferred $(C/A,B/A)$ distribution (second row). 
Further fixing triaxiality-related parameters (rows 3--4) produces comparatively smaller changes, particularly in bins with larger sample sizes, consistent with the interpretation that the dominant limitation is the weak identifiability of triaxiality parameters from projected data alone.

Overall, both models recover a qualitatively prolate-dominated population with a spheroidal component; however, $\ETa$ tends to shift probability toward the boundaries between the shape classes, whereas $\ltETa$ naturally reproduces the asymmetric/skewed intrinsic distributions measured from particles (Fig.~\ref{measured 3D shapes}) because Gaussianity is imposed in $(\tilde{E},\tilde{T})$ rather than in $(E,T)$.

\section{Binney Projection Formalism}
\label{Appendix E}
\begin{figure*}
    \centering
    \includegraphics[width=\textwidth]{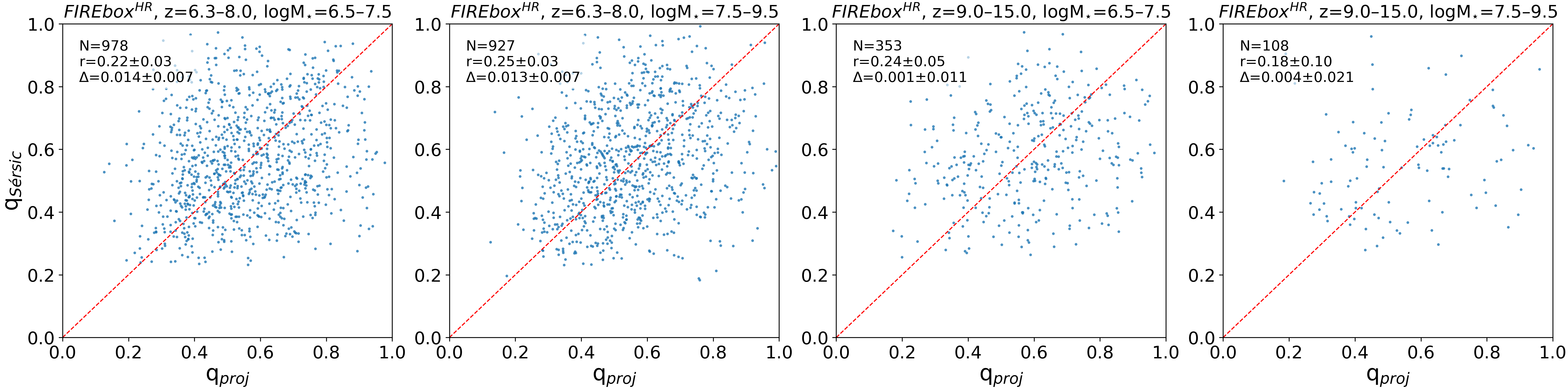}
    \includegraphics[width=\textwidth]{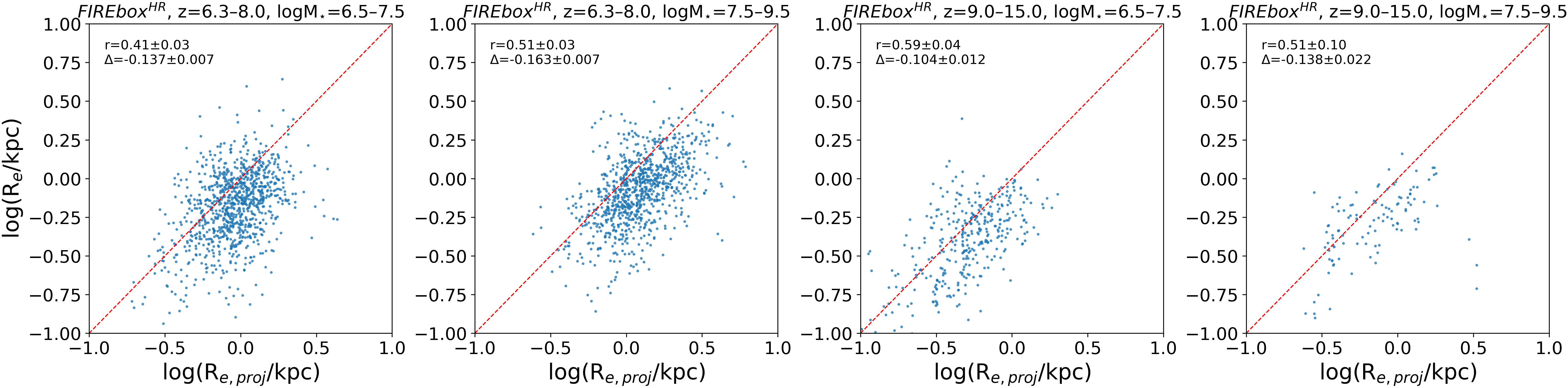}
    \includegraphics[width=\textwidth]{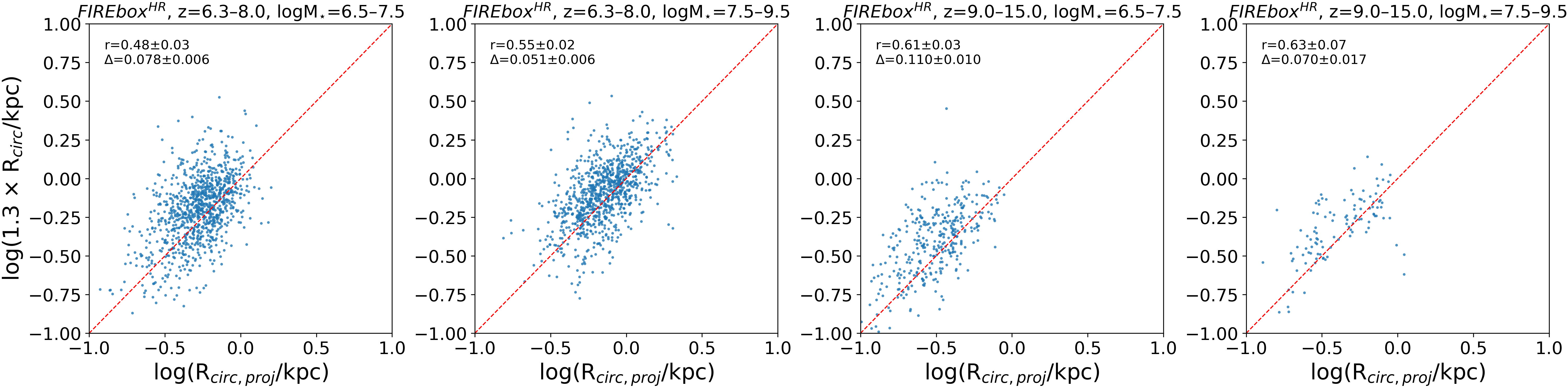}
    \caption{Comparison of 2D structural parameters derived from S\'ersic fitting of mock observations ($y$-axis) versus those predicted by projecting 3D MCD-measured shapes using the Binney formalism ($x$-axis).
    The columns correspond to different redshift and mass bins in the \FIRE simulation.
    \textbf{Top Row:} Comparison of projected axis ratios ($q=b/a$).
    \textbf{Middle Row:} Comparison of semi-major axes ($\log R_e$). A systematic bias is observed due to differing size definitions (half-light vs. half-mass).
    \textbf{Bottom Row:} Comparison of circularized radii ($\log R_{\text{circ}}$), where the S\'ersic radius is scaled by a factor of $1.3$ to match the theoretical half-mass definition \citep{vandeVen2021}.
    The Pearson correlation coefficient ($r$) and mean bias ($\Delta$) are indicated in each panel. While axis ratios show reasonable agreement, a residual offset remains in the size measurements even after circularization corrections.}
    \label{Fig E: Projection vs. Sersic}
\end{figure*}

In this work, we use the Binney \citep{Binney1985} projection formalism, which treats galaxies as triaxial solid ellipsoids that project onto two-dimensional ellipses. 
The projected size is defined as the semi-major axis, and the axis ratio $b/a$ is the ratio of the semi-minor to the semi-major axis.
\citet{Zhang2019} (Appendix D) discuss the differences between the 2D structure projected via the Binney formalism and 2D S\'ersic profiles produced by projecting a 3D profile \citep{vandeVen2021}. 
Although qualitatively similar, the latter is likely a more realistic approach as it reproduces 2D S\'ersic profiles exactly.
In our HMC inference model, a more influential factor is the discrepancy in size definitions: our theoretical galaxy size is defined as the radius enclosing half the mass within an ellipsoid characterized by MCD-measured 3D shapes.
The projected half-mass radius is expected to differ from the half-light radius as they are intrinsically distinct quantities with no trivial conversion. 
However, it has been suggested that for specific galaxy populations, the spherically measured half-mass radius is an unbiased proxy for 1.3 times the circularized half-light radius \citep{vandeVen2021}. We examine each of these components below.

Figure \ref{Fig E: Projection vs. Sersic} presents scatter plots comparing the S\'ersic-measured 2D structural parameters $(q_{\text{S\'ersic}}\equiv b/a, \log(R_e/\text{kpc}))$ with the projected 2D parameters $(q_{\text{proj}}, \log(R_{e,\text{proj}}/\text{kpc}))$ derived via the Binney formalism from MCD-measured 3D shapes.
The first row compares the axis ratios. The upper-left corner of each panel displays the Pearson correlation coefficient, $r$, and the mean bias, $\Delta\equiv \text{mean}(y-x)$. 
The correlation between $q_{\text{S\'ersic}}$ and $q_{\text{proj}}$ is relatively weak, likely because the galaxies are viewed at random angles and are prolate-dominated, spanning a wide range of projected shapes. The bias is small, however, indicating good agreement between the axis ratios derived from the two methods. 
For the galaxy sizes presented in the second row, a clear bias exists, primarily due to the discrepancy in size definitions. 
To resolve this tension, we circularize both the ``1.3 times half-light radius'' and the ``projected half-mass radius'', as discussed above. 
The bias decreases but remains non-negligible. 
Moreover, the correlations between $b/a$ and size within the joint distribution $(b/a, \log a)$ differ by a small amount ($\sim 0.15$).
These observations suggest that while our model applied to the \FIRE samples is not perfectly comparable, the ``true'' 3D shapes are still recovered with reasonable accuracy. 
For a more consistent test, future work should consider measuring the ``true'' shape using a light-weighted estimator, such as that used in \citet{klein2026}. 
Furthermore, by generating an ellipsoid library using light-based 3D profiles that strictly recover 2D S\'ersic profiles upon projection, we can develop a complete Bayesian inference method to accurately recover the 3D geometry of galaxies.

%%%%%%%%%%%%%%%%%%%%%%%%%%%%%%%%%%%%%%%%%%%%%%%%%%%%%%%%%%%%%%%
% In the PDF output, floats should be placed
% under their own appendix, not before the title, nor after the
% title of the next appendix.

% In short appendices, onecolumn floats (\figure*
% or \table*) will generate a blank page.
% To prevent this behaviour, a few examples are provided here. 

% In case you have a lot of floating objects for little text and the 
% LaTeX engine moves the floats away from their context, the command
% \FloatBarrier of the “placeins” package will empty the
% float buffer and place all stored floats in the continuity.

% If you still encounter problems with wide floats placement,
% just use the onecolumn environment throughout the appendices.
%%%%%%%%%%%%%%%%%%%%%%%%%%%%%%%%%%%%%%%%%%%%%%%%%%%%%%%%%%%%%%%
\end{appendix}
\end{document}